\documentclass[12pt]{article}
\pdfoutput=1
\usepackage[utf8]{inputenc}
\usepackage[a4paper, total={16.5cm, 22.5cm}]{geometry}
\usepackage{hyperref} 
\usepackage{amsmath}
\usepackage{amssymb}
\usepackage{subfig} 
\usepackage{physics} 
\usepackage{soul}
\usepackage{epsfig}
\usepackage{cite}

\def\bc{\begin{center}}
\def\ec{\end{center}}
\def\beq{\begin{equation}}
\def\eeq{\end{equation}}
\def\beqn{\begin{eqnarray}}
\def\eeqn{\end{eqnarray}}
\def\no{\nonumber}
\def\eqn#1{(\ref{#1})}
\def\ba{\begin{array}{c}}
\def\ea{\end{array}}
\def\bat{\begin{array}{cc}}
\def\bi{\begin{itemize}}
\def\ei{\end{itemize}}

\def\cL{{\cal L}}

\def\cC{{\cal C}}
\def\cD{{\cal D}}

\def\cP{{\cal P}}
\def\cH{{\cal H}}
\def\cO{{\cal O}}

\def\cR{{\cal R}}

\def\cZ{{\cal Z}}

\newcommand{\lsim}{~{}_{\textstyle\sim}^{\textstyle <}~}

\newcommand{\e}{\mathrm{e}}

\usepackage{tikz}
\usetikzlibrary{"arrows", "automata", "backgrounds", "calendar", "chains", "matrix", "mindmap", "patterns", "petri", "shadows", "shapes.geometric", "shapes.misc", "spy", "trees"}
\graphicspath{{./images/}}
\usepackage{appendix}

\usepackage{tikz}
\usetikzlibrary{arrows,shapes}
\usetikzlibrary{trees}
\usetikzlibrary{matrix,arrows} 				
\usetikzlibrary{positioning}				
\usetikzlibrary{calc,through}				
\usetikzlibrary{decorations.pathreplacing}  
\usepackage{pgffor}							

\usetikzlibrary{decorations.pathmorphing}	
\usetikzlibrary{decorations.markings}
\tikzset{
    vector/.style={decorate, decoration={snake}, draw},
	provector/.style={decorate, decoration={snake,amplitude=2.5pt}, draw},
	antivector/.style={decorate, decoration={snake,amplitude=-2.5pt}, draw},
    fermion/.style={draw=black, postaction={decorate},
        decoration={markings,mark=at position .55 with {\arrow[draw=black]{>}}}},
    fermionbar/.style={draw=black, postaction={decorate},
        decoration={markings,mark=at position .55 with {\arrow[draw=black]{<}}}},
    fermionnoarrow/.style={draw=black},
    gluon/.style={decorate, draw=black,
        decoration={coil,amplitude=4pt, segment length=5pt}},
    scalar/.style={dashed,draw=black, postaction={decorate},
        decoration={markings,mark=at position .55 with {\arrow[draw=black]{>}}}},
    scalarbar/.style={dashed,draw=black, postaction={decorate},
        decoration={markings,mark=at position .55 with {\arrow[draw=black]{<}}}},
    scalarnoarrow/.style={dashed,draw=black},
    electron/.style={draw=black, postaction={decorate},
        decoration={markings,mark=at position .55 with {\arrow[draw=black]{>}}}},
	bigvector/.style={decorate, decoration={snake,amplitude=4pt}, draw},
}

\tikzset{cross/.style={cross out, draw=black, minimum size=2*(#1-\pgflinewidth), inner sep=0pt, outer sep=0pt},
cross/.default={1pt}}
\tikzstyle{block} = [draw, rectangle, 
    minimum height=3em, minimum width=6em]

\usepackage{xparse}
\NewDocumentCommand\semiloop{O{black}mmmO{}O{above}}
{%
\draw[#1] let \p1 = ($(#3)-(#2)$) in (#3) arc (#4:({#4+180}):({0.5*veclen(\x1,\y1)})node[midway, #6] {#5};)
}

\begin{document}
 \pdfoutput=1

\thispagestyle{empty}
\hfill\mbox{IFIC/17-32, FTUV/17-1005}
\vspace{3cm}

\begin{center}

{\huge\sc \bf
Flavour alignment in\\[10pt] multi-Higgs-doublet models
}

\vspace{1cm}

{\sc
Ana Peñuelas and Antonio Pich
}

\vspace*{.7cm}

{\sl
Departament de F\'\i sica Te\`orica, IFIC, Universitat de Val\`encia -- CSIC,\\
Apt. Correus 22085, E-46071 Val\`encia, Spain,
}

\end{center}

\vspace*{0.1cm}

\begin{abstract}
\noindent
Extended electroweak scalar sectors containing several doublet multiplets require
flavour-aligned Yukawa matrices to prevent the appearance at tree level of 
unwanted flavour-changing neutral-current transitions. We analyse the misalignment induced by one-loop quantum corrections and explore possible generalizations of the alignment condition and their compatibility with current experimental constraints.
The hypothesis of flavour alignment at a high scale turns out to be consistent with all known phenomenological tests.

\end{abstract}

\vfill\hfill\mbox{} 

\setcounter{page}{0}

\newpage

\section{Introduction}
\label{sec:intro}

Scalar multiplets transforming as doublets or singlets under the $SU(2)_L$ gauge group are the favoured candidates for building extended models of perturbative electroweak symmetry breaking (EWSB), beyond the Standard Model (SM) framework \cite{Pich:2015lkh}. Assigning a zero hypercharge to the singlets and $Y=Q-T_3=\frac{1}{2}$ to the doublet scalars, these models automatically satisfy the successful mass relation $M_W = M_Z\,\cos{\theta_W}$ and can then be easily adjusted to fulfil all precision electroweak tests. The observable signals of the singlet scalar fields are quite restricted because they
do not have Yukawa interactions with the SM fermions, nor they couple to the gauge bosons.
Therefore, they can only communicate with those SM particles through their mixing with other neutral scalars in non-singlet multiplets. 

Doublet fields give rise to a much more interesting phenomenology with non-trivial implications for the fermionic flavour dynamics. In addition to the three electroweak Goldstone bosons, the spectrum of a scalar sector composed by $N$ doublets
contains $N-1$ charged fields $H^\pm$ and $2N-1$ neutral scalars, with a rich variety of possible interactions. In general, these include Yukawa couplings of the neutral scalars that are not diagonal in flavour, implying dangerous flavour-changing neutral-current (FCNC) transitions, which are tightly constrained experimentally \cite{Pich:2011nh}. 

To avoid the presence of unwanted FCNC phenomena, one must impose ad-hoc dynamical restrictions, suppressing these effects below the empirically forbidden level. The models most frequently considered in the literature \cite{Branco:2011iw,Gunion:1989we,Ivanov:2017dad} assume that only one single scalar doublet can couple to a given type of right-handed fermion $f_R$. This guarantees identical flavour structures for the Yukawa interactions and the fermion mass matrices, so that FCNC vertices are absent as in the SM. While this assumption is quite strong, it can be easily implemented in the models, enforcing appropriately defined discrete $\cZ_2$ symmetries which forbid the Yukawa couplings of all other scalar doublets to $f_R$ \cite{Haber:1978jt,Hall:1981bc,Donoghue:1978cj,Barger:1989fj,Deshpande:1977rw,Grossman:1994jb,Akeroyd:1994ga,Akeroyd:1996he,Ma:2008uza,Aoki:2009ha} and keep the resulting flavour structure stable under quantum corrections (natural flavour conservation) \cite{Glashow:1976nt,Paschos:1976ay}. 

Flavour alignment \cite{Pich:2009sp,Pich:2010ic} is a much more general possibility, based on the weaker assumption that the couplings of all scalar doublets to a given right-handed fermion have the same flavour structure \cite{Pich:2009sp,Pich:2010ic,Manohar:2006ga}. All Yukawas can then be diagonalized simultaneously, eliminating the FCNC vertices from the tree-level Lagrangian. FCNCs effects reappear at higher perturbative orders because quantum corrections misalign the different Yukawas \cite{Ferreira:2010xe,Jung:2010ik,Botella:2015yfa}. However, the build-in flavour symmetries strongly constrain the possible FCNC operators that can be generated at the quantum level \cite{Pich:2009sp,Pich:2010ic}, implying an effective theory with minimal flavour violation \cite{Chivukula:1987py,DAmbrosio:2002vsn}.

The induced one-loop FCNC Yukawas have been explicitly analysed within the aligned two-Higgs-doublet model (A2HDM) \cite{Pich:2009sp,Pich:2010ic,Jung:2010ik,Braeuninger:2010td,Bijnens:2011gd,Li:2014fea,Abbas:2015cua,Gori:2017qwg}, and their effects have been found to be small and well below all known experimental constraints, giving further support to the successful phenomenology of this particular new-physics scenario \cite{Jung:2010ik,Li:2014fea,Abbas:2015cua,Gori:2017qwg,Jung:2010ab,Jung:2012vu,Celis:2013rcs,Celis:2013ixa,Ilisie:2014hea,Jung:2013hka,Ilisie:2015tra,Cherchiglia:2016eui,Chang:2015rva,Hu:2016gpe,Hu:2017qxj,Cho:2017jym,Altmannshofer:2012ar,Bai:2012ex,Duarte:2013zfa,Ayala:2016djv,Han:2015yys,Enomoto:2015wbn,Mileo:2014pda,Wang:2013sha,Akeroyd:2012yg,Cree:2011uy,Serodio:2011hg,Lopez-Val:2013yba}. However, some recent flavour anomalies observed in $B\to D^{(*)}\tau\nu$ data \cite{Lees:2012xj,Lees:2013uzd,Huschle:2015rga,Sato:2016svk,Hirose:2016wfn,Hirose:2017dxl,Aaij:2015yra,Aaij:2017uff}
have triggered the consideration of flavour non-universal aligned-like structures \cite{Mahmoudi:2009zx,Celis:2012dk,Celis:2016azn,Crivellin:2013wna,Cline:2015lqp,Buras:2010mh,Dery:2013aba,Ahn:2010zza, Crivellin:2012ye, Crivellin:2015hha}, which have not been explored at the quantum level.

In the following, we present a detailed study of the stability of flavour alignment under quantum corrections. We analyse the FCNC operators generated at one loop for a generic scalar sector with $N$ doublets, both for the flavour-aligned model and for its generalization with non-universal aligned-like structures. We want to understand the quantum structure of these models and their phenomenological viability. We discuss first in section~\ref{sec:multiHiggs} the general Yukava Lagrangian of the N-Higgs-doublet model,
and briefly describe in section~\ref{sec:NFC} the usual models with natural flavour conservation. The alignment assumption is implemented in section~\ref{sec:alignment}, where its possible generalizations are discussed. The one-loop renormalization-group equations (RGEs) of the model are used in section~\ref{sec:rge} to pin down the induced FCNC operators in the most general case. The result is then particularized to the different situations we are interested in, and the usual scenarios with $\cZ_2$ symmetries are easily recovered. Section~\ref{sec:symmetry} analyses the underlying symmetries governing the specific flavour structures obtained through the RGEs.
The phenomenological implications are discussed in sections~\ref{sec:phenomenology}, \ref{sec:Bs} and \ref{sec:mesonmixing}, and a brief summary is finally given in section~\ref{sec:summary}. Some technical details are relegated to the appendix.

\section{Multi-Higgs-doublet models}
\label{sec:multiHiggs}

Let us consider an electroweak model with the SM fermion content and gauge group, and an extended scalar sector involving $N$ doublets with hypercharge $Y=\frac{1}{2}$, 
\beq
\phi_a\, =\, \e^{i\theta_a}\;\left[\ba \phi_a^+\\ \frac{1}{\sqrt{2}}\, (v_a+\rho_a+i\,\eta_a)\ea\right]\, .
\eeq
%
Their neutral components acquire vacuum expectation values $\langle\phi_a^0\rangle =\e^{i\theta_a}\, v_a/\sqrt{2}$, which in full generality could be complex ($v_a\ge 0$). One global phase can always be rotated away through a $U(1)_Y$ transformation; we choose $\theta_1=0$, leaving the relative phases $\tilde\theta_a = \theta_a - \theta_1$.
For our discussion, it is not necessary to specify the scalar potential and gauge couplings. We are only interested in the Yukawa interactions which take the generic form
\beq\label{eq:Yukawa}
\cL_Y\, =\, -\sum_{a=1}^N\,\left\{\bar Q_L' \left( \Gamma_a\phi_a\, d_R' + \Delta_a\tilde\phi_a\, u_R'\right) + \bar L_L'\, \Pi_a\phi_a\, \ell_R' + \mathrm{h.c.}\right\}\, ,
\eeq
where $\tilde\phi_a \equiv i\tau_2\phi_a^*$ are the charge-conjugated scalar fields, $Q_L'$ and $L_L'$ the left-handed quark and lepton doublets, and $d_R'$, $u_R'$, $\ell_R'$ the corresponding right-handed fermion singlets. All fermion fields denote $N_G=3$ vectors
in flavour space; for instance, $d_R' = (d'_R, s'_R, b'_R)^T$. The Yukawa couplings $\Gamma_a$, $\Delta_a$ and $\Pi_a$ are $N_G\times N_G$ complex flavour matrices.

It is convenient to perform a global $SU(N)$ transformation in the space of scalar fields,
\beq
\Phi_a\, =\, \sum_{b=1}^N \Omega_{ab}\,\e^{-i\tilde\theta_b}\,\phi_b\, ,
\qquad\qquad
\phi_b\, =\,\e^{i\tilde\theta_b}\,\sum_{a=1}^N \Omega_{ab}\,\Phi_a\, ,
\qquad\qquad \Omega\cdot\Omega^T\, =\,\Omega^T\cdot\Omega\, =\, 1\, ,
\eeq
such that only the first doublet acquires a vacuum expectation value. The needed transformation is characterized by the condition $\Omega_{1a} = v_a/v$, with
$v = \left( \sum_a v_a^2\right)^{1/2} > 0$, and defines the Higgs basis
\beq
\Phi_1\, =\,\left[\ba G^+\\ \frac{1}{\sqrt{2}}\, (v+S_1^0+i\, G^0)\ea\right]\, ,
\qquad\qquad
\Phi_{a>1}\, =\,\left[\ba S_a^+\\ \frac{1}{\sqrt{2}}\, (S_a^0+i\, P_a^0)\ea\right]\, .
\eeq
The EWSB is then fully associated with the doublet $\Phi_1$, which incorporates the 
electroweak Goldstone fields $G^0$ and $G^+$, and plays the role of the SM Higgs doublet. 

The physical mass-eigenstate charged (neutral) scalars are linear combinations of the $S_a^+$ ($S_a^0$ and $P_a^0$) fields.
The $2N-1$ neutral scalar mass eigenstates, $\varphi_i^0 = \mathcal{R}_{ij} \mathcal{S}^0_j$, are related with the scalar-doublet field components  
$\mathcal{S}^0_i = \{ S^0_1, S^0_2, P^0_2, \cdots , S^0_N, P^0_N \}$
through an orthogonal transformation $\mathcal{R}$ which depends on the parameters of the scalar potential. With a CP-conserving potential, the neutral scalar mixing matrix splits into two separate CP-even ($S^0_a$) and CP-odd ($P^0_a$) mixing structures. CP-violation mixes the two scalar sectors and the resulting mass eigenstates do not have, in general, definite CP quantum numbers.
Similarly, the $N-1$ charged fields $\mathcal{S}^+_i = \{ S^+_2, S^+_3, \cdots , S^+_N\}$ mix among themselves giving rise to the charged mass eigenstates 
$\varphi_i^+ = \mathcal{R}_{ij}^{(+)} \mathcal{S}^+_j$, with $\mathcal{R}^{(+)}$ a $(N-1)\times(N-1)$ orthogonal matrix.

In the Higgs basis the Yukawa structures in Eq.~\eqn{eq:Yukawa} take the form
\beq
\sum_{a=1}^N\,\Gamma_a\phi_a\, = \,\sum_{b=1}^N\,\hat\Gamma_b\Phi_b\, ,
\qquad\qquad
\sum_{a=1}^N\,\Delta_a\tilde\phi_a\, = \,\sum_{b=1}^N\,\hat\Delta_b\tilde\Phi_b\, ,
\qquad\qquad
\sum_{a=1}^N\,\Pi_a\phi_a\, = \,\sum_{b=1}^N\,\hat\Pi_b\Phi_b\, ,
\eeq
with
\beq
\hat\Gamma_b\, =\,\sum_{a=1}^N \Omega_{ba}\,\e^{i\tilde\theta_a}\,\Gamma_a\, ,
\qquad\qquad
\hat\Delta_b\, =\,\sum_{a=1}^N \Omega_{ba}\,\e^{-i\tilde\theta_a}\,\Delta_a\, ,
\qquad\qquad
\hat\Pi_b\, =\,\sum_{a=1}^N \Omega_{ba}\,\e^{i\tilde\theta_a}\,\Pi_a\, .
\eeq

The EWSB mechanism generates the mass matrices
\beq
M'_d\, =\, \frac{v}{\sqrt{2}}\, \hat\Gamma_1\, ,
\qquad\qquad
M'_u\, =\, \frac{v}{\sqrt{2}}\, \hat\Delta_1\, ,
\qquad\qquad
M'_\ell\, =\, \frac{v}{\sqrt{2}}\, \hat\Pi_1\, ,
\eeq
which only involve the Yukawa structures associated with the doublet field $\Phi_1$.
Their diagonalization determines the fermion mass eigenstates
\beq
U^{f\dagger}_L\, M'_f\, U^f_R\, =\, M_f\, ,
\qquad\qquad
f'_L\, =\, U^f_L\, f_L\, ,
\qquad\qquad
f'_R\, =\, U^f_R\, f_R\, ,
\eeq
and the fermion masses 
\beq
M_d\, =\, \mathrm{diag} (m_d, m_s, m_b)\, ,
\qquad\quad
M_u\, =\, \mathrm{diag} (m_u, m_c, m_t)\, ,
\qquad\quad
M_\ell\, =\, \mathrm{diag} (m_e, m_\mu, m_\tau)\, .
\eeq
Neutrinos remain massless because the model does not include $\nu_R$ fields.

In terms of the fermion mass eigenstates, the Yukawa Lagrangian is given by
\beqn\label{eq:YL}
\cL_Y & = & - \left( 1 + \frac{S_1^0}{v}\right) \,
\left\{ \bar d_L M_d d_R + \bar u_L M_u u_R + \bar \ell_L M_\ell \ell_R \right\}
\no\\
&&\mbox{} -\frac{1}{v} \;\sum_{a=2}^N\, \left( S_a^0 + i\, P_a^0\right)
\left\{  \bar d_L Y_d^{(a)} d_R + \bar u_R Y_u^{(a)\dagger} u_L + \bar \ell_L Y_\ell^{(a)} \ell_R\right\}
\no\\
&&\mbox{} -\frac{\sqrt{2}}{v}\;\sum_{a=2}^N\, S^+_a \left\{ 
\bar u_L V_{_{\mathrm{CKM}}} Y^{(a)}_d d_R - \bar u_R Y_u^{(a)\dagger} V_{_{\mathrm{CKM}}} d_L  + \bar\nu_L Y^{(a)}_\ell \ell_R\right\} 
+ \mathrm{h.c.}\, ,
\eeqn
where $V_{_{\mathrm{CKM}}} = U^{u\dagger}_L U^d_L$ is the usual Cabibbo-Kobayashi-Maskawa (CKM) quark-mixing matrix \cite{Cabibbo:1963yz, Kobayashi:1973fv}.
The analogous mixing matrix  in the charged-current leptonic Yukawa, $V_{_{\mathrm{L}}} = U^{\nu\dagger}_L U^\ell_L$, has been reabsorbed through a redefinition of the massless neutrino fields, $\bar\nu_L\cdot V_{_{\mathrm{L}}}\to \bar\nu_L$, so that the leptonic $W^\pm$ interactions are flavour diagonal.
For $a\not=1$, the Yukawa structures
\beq
Y^{(a)}_d\, =\, \frac{v}{\sqrt{2}}\; U^{d\dagger}_L\,\hat\Gamma_a\, U^d_R\, ,
\qquad
Y^{(a)}_u\, =\, \frac{v}{\sqrt{2}}\; U^{u\dagger}_L\,\hat\Delta_a\, U^u_R\, ,
\qquad
Y^{(a)}_\ell\, =\, \frac{v}{\sqrt{2}}\; U^{\ell\dagger}_L\,\hat\Pi_a\, U^\ell_R \, , 
\eeq
are not related to the mass matrices and their elements could take arbitrary complex values. 
In general, they remain non-diagonal in the fermion mass-eigenstate basis, giving rise to unwanted flavour-changing couplings of the neutral scalar fields.

\section{Natural flavour conservation}
\label{sec:NFC}

The simplest way to avoid flavour non-diagonal Yukawa matrices $Y_f^{(a)}$ is minimizing drastically the number of flavour structures in the Lagrangian \eqn{eq:Yukawa} so that, for a given type of right-handed fermion $f'_R$, only one single scalar doublet $\phi_{a_f}$ is allowed to have non-zero Yukawa coupling. A given choice of three fields $\{\phi_{a_d},\phi_{a_u},\phi_{a_\ell}\}$ defines a particular model with
$\Gamma_a = \delta_{a_d a}\, \Gamma_{a_d}$, $\Delta_a = \delta_{a_u a}\, \Delta_{a_u}$ and $\Pi_a = \delta_{a_\ell a}\, \Pi_{a_\ell}$.

In the Higgs basis, this implies
\beq
\hat \Gamma_a\, =\, \Omega_{a a_d}\, \e^{i\tilde\theta_{a_d}}\,\Gamma_{a_d}\, ,
\qquad
\hat \Delta_a\, =\, \Omega_{a a_u}\, \e^{-i\tilde\theta_{a_u}}\,\Delta_{a_u}\, ,
\qquad
\hat \Pi_a\, =\, \Omega_{a a_\ell}\, \e^{i\tilde\theta_{a_\ell}}\,\Pi_{a_\ell}\, .
\eeq
Since there are only three flavour structures, one for each type of fermion, the diagonalization of the mass matrices $\hat\Gamma_1$, $\hat\Delta_1$ and $\hat\Pi_1$ also diagonalizes all Yukawas with $a\not=1$ \cite{Pich:2009sp, Pich:2010ic}. One obtains:
\beq
Y_f^{(a)}\, =\, \varsigma_f^{(a)} \; M_f\, ,
\qquad\qquad
\varsigma_f^{(a)}\, =\,\frac{\Omega_{a a_f}}{\Omega_{1 a_f}}\, .
\eeq

This particular form of the Yukawa Lagrangian could be enforced through a discrete symmetry 
$\cZ_2^d\otimes\cZ_2^u\otimes\cZ_2^\ell$, where each separate $\cZ_2^f$ transformation is defined so that $f'_R$ and $\phi_{a_f}$ reverse sign,
\beq
\cZ_2^f :\quad f'_R \,\to\, - f'_R\, ,
\qquad\qquad
\phi_{a_f} \, \to\, -  \phi_{a_f}\, ,
\eeq
while all other fields remain unchanged \cite{Weinberg:1976hu}. 
The symmetry guarantees that the resulting flavour structure is stable under quantum corrections, ensuring that FCNC local interactions cannot reappear at higher orders.
Notice that the assumption of natural flavour conservation singles out a particular basis of scalar fields where the discrete symmetry is defined.

For $N=2$, one can choose four different inequivalent options for $\{ a_d, a_u, a_\ell\}$, where $a_f$ labels the doublet to which the fermion $f'_R$ is coupled
(the remaining possibilities amount to a permutation of $\phi_1$ and $\phi_2$), which are usually taken as
\beq
\begin{array}{llc}
\mathrm{Type\; I:} & \{ 2, 2, 2\}\, ,
\qquad &
\varsigma_d = \varsigma_u = \varsigma_\ell = \cot{\beta}\, ,
\\
\mathrm{Type\; II:} & \{ 1, 2, 1\}\, , & 
\varsigma_d = \varsigma_\ell = -\tan{\beta}\, ,
\qquad\varsigma_u = \cot{\beta}\, ,
\\
\mathrm{Type\; X:} & \{ 2, 2, 1\}\, , &
\varsigma_d = \varsigma_u = \cot{\beta}\, ,
\qquad \varsigma_\ell = -\tan{\beta}\, ,
\\
\mathrm{Type\; Y:} & \{ 1, 2, 2\}\, , &
\varsigma_d = -\tan{\beta}\, , \qquad
\varsigma_u = \varsigma_\ell = \cot{\beta}\, ,
\ea
\eeq
with $\varsigma_f\equiv \varsigma_f^{(2)}$ and $\tan{\beta}\equiv v_2/v_1$. 
A single $\cZ_2$ transformation is enough in this case to define the model: $\phi_1$ is odd, while $\phi_2$, $Q'_L$, $L'_L$ and $u'_R$ are all even. The four different types of models are obtained defining different transformations of the $d'_R$ and $\ell'_R$ fields under $\cZ_2$. In type I the two fields are even \cite{Haber:1978jt, Hall:1981bc}, they are both odd in type II \cite{ Hall:1981bc,  Donoghue:1978cj}  $d'_R\to d'_R$ and $\ell'_R\to - \ell'_R$ in type X \cite{ Barger:1989fj}, and $d'_R\to - d'_R$ and $\ell'_R\to \ell'_R$ in type Y \cite{ Barger:1989fj }.
If the $\cZ_2$ symmetry is imposed in the Higgs basis, all fermions must couple to $\Phi_1$ in order to get their masses and the doublet $\Phi_2$ necessarily decouples from the fermion sector. One gets then a type-I structure (exchanging the labels 1 and 2) with $\varsigma_f = 0$, known as the inert two-Higgs-doublet model \cite{Deshpande:1977rw}.

With $N=3$ there are five inequivalent possibilities, up to permutations of the three scalar-field labels, which we define through the following choices of $\{ a_d, a_u, a_\ell\}$:
\beq
\begin{array}{llc}
\mathrm{Type\; A:} & \{ 1, 1, 1\}\, ,
\qquad &
\varsigma_d^{(a)} = \varsigma_u^{(a)} = \varsigma_\ell^{(a)} = \Omega_{a1}/\Omega_{11}
\\
\mathrm{Type\; B:} & \{ 1, 2, 1\}\, , & 
\varsigma_d^{(a)} = \varsigma_\ell^{(a)} = \Omega_{a1}/\Omega_{11}\, ,
\qquad\varsigma_u^{(a)} =\Omega_{a2}/\Omega_{12}\, ,
\\
\mathrm{Type\; C:} & \{ 1, 1, 2\}\, , &
\varsigma_d^{(a)} = \varsigma_u^{(a)} = \Omega_{a1}/\Omega_{11}\, ,
\qquad \varsigma_\ell^{(a)} = \Omega_{a2}/\Omega_{12}\, ,
\\
\mathrm{Type\; D:} & \{ 1, 2, 2\}\, , &
\varsigma_d^{(a)} = \Omega_{a1}/\Omega_{11}\, , \qquad
\varsigma_u^{(a)} = \varsigma_\ell^{(a)} = \Omega_{a2}/\Omega_{12}\, ,
\\
\mathrm{Type\; E:} & \{ 1, 2, 3\}\, . &
\varsigma_d^{(a)} = \Omega_{a1}/\Omega_{11}\, , \qquad
\varsigma_u^{(a)} = \Omega_{a2}/\Omega_{12}\, , \qquad
\varsigma_\ell^{(a)} = \Omega_{a3}/\Omega_{13}\, .
\ea
\eeq
One can easily check that each one of these structures can be enforced by using only two $\cZ_2$ symmetries.

For $N> 3$, natural flavour conservation implies that three scalar doublets, which can always be chosen as $\phi_{1,2,3}$, couple to the fermions following one of the five allowed $N=3$ types, while the remaining $N-3$ doublets decouple.

\section{Flavour alignment}
\label{sec:alignment}

Natural flavour conservation is a very strong assumption, which for $N>3$ involves $N-3$ fermiophobic scalar doublets (in the scalar basis where the $\cZ_2^f$ symmetries are imposed).  
In order to avoid FCNC interacting vertices in $\cL_Y$, what is really needed is that only a single flavour structure is present for each $f_R$ type, {\it i.e.}, the alignment condition \cite{Pich:2009sp,Pich:2010ic}:
\beq
\Gamma_a\, =\, \e^{-i\tilde\theta_a}\,\xi^{(a)}_d\; \Gamma_1\, , 
\qquad\quad
\Delta_a \, =\, \e^{i\tilde\theta_a}\,\xi^{(a)\dagger}_u\; \Delta_1\, ,
\qquad\quad
\Pi_a \, =\, \e^{-i\tilde\theta_a}\,\xi^{(a)}_\ell\; \Pi_1 \, ,
\eeq
where $\xi^{(1)}_f = 1$ while $\xi^{(a\not= 1)}_f$ can be arbitrary complex parameters. All Yukawa matrices are then simultaneously diagonalized in the fermion mass-eigenstate basis, with the result
\beq\label{eq:alignment}
Y_{d,\ell}^{(a)}\, =\, \varsigma_{d,\ell}^{(a)} \; M_{d,\ell}\, ,
\qquad\qquad\qquad
Y_u^{(a)}\, =\, \varsigma_u^{(a)\dagger} \; M_u\, ,
\eeq
where the alignment proportionality parameters are given by
\beq
\varsigma_f^{(a)}\, =\,\frac{\sum_{b=1}^N \Omega_{a b}\,\xi_f^{(b)}}{\sum_{b=1}^N \Omega_{1 b}\,\xi_f^{(b)}}\, .
\eeq
Natural flavour conservation corresponds to the particular cases where the alignment parameters $\xi_f^{(b\not = 1)}$ are either all zero ($\varsigma_f^{(a)} = \Omega_{a1}/\Omega_{11}$) or one of them, $\xi_f^{(a_f)}$, takes an infinite value  ($\varsigma_f^{(a)} = \Omega_{aa_f}/\Omega_{1a_f}$).

The hypothesis of flavour alignment leads to a very appealing structure for the Yukawa Lagrangian in Eq.~\eqn{eq:YL}: i) all fermion-scalar interactions are proportional to the corresponding fermion mass matrices, ii) FCNCs vertices are absent at tree level, and iii) the only source of flavour-changing transitions is the charged-current quark mixing matrix $V_{_\mathrm{CKM}}$, which appears in the $W^\pm$ and $H^\pm$ fermionic couplings. In addition to the fermion masses, the only new parameters introduced by the Yukawa interactions are the $3 (N-1)$ complex alignment factors $ \varsigma_f^{(a)}$ ($a\not= 1$), which provide additional sources of CP violation beyond the SM quark-mixing phase.

The flavour-alignment condition does not exhaust all possibilities for a tree-level Lagrangian without FCNC interactions. The most general structure is obtained with a set of $N$ simultaneously-diagonalizable matrices $Y_f^{\prime (a)}$, for each type of fermion $f$. 
One can also describe this generic possibility with the
parametrization \eqn{eq:alignment} through the alignment matrices 
\beq\label{eq:GAlignment}
\varsigma_{d,\ell}^{(a)} \,\equiv\, Y_{d,\ell}^{(a)}\, M_{d,\ell}^{-1}\, ,
\qquad\qquad\quad
\varsigma_u^{(a)\dagger} \,\equiv\, Y_u^{(a)}\, M_u^{-1}\, .
\eeq
These expressions are completely general because all charged fermion masses are known to be non vanishing; therefore, $\det M_f\not= 0$ and $M_f^{-1}$ is well defined. Since all $Y_f^{(a)}$ matrices are assumed to be diagonal, the alignment factors become now diagonal matrices (in the fermion mass-eigenstate basis):
\beq
\varsigma_d^{(a)}\, =\, \mathrm{diag} (\varsigma_d^{(a)}, \varsigma_s^{(a)}, \varsigma_b^{(a)})\, ,
\qquad
\varsigma_u^{(a)}\, =\, \mathrm{diag} (\varsigma_u^{(a)}, \varsigma_c^{(a)}, \varsigma_t^{(a)})\, ,
\qquad
\varsigma_\ell^{(a)}\, =\, \mathrm{diag} (\varsigma_e^{(a)}, \varsigma_\mu^{(a)}, \varsigma_\tau^{(a)})\, .
\eeq
The structure of the resulting Yukawa Lagrangian in Eq.~\eqn{eq:YL} is formally the same than for normal alignment (provided one takes care of not commuting the matrix factors
$\varsigma_f^{(a)}$ and $V_{_\mathrm{CKM}}$). However, one loses the hierarchies dictated by the fermion mass spectrum because there is really no connection between the numerical values of the Yukawa couplings and the corresponding masses. Small (large) values of $m_f$ can be compensated with large (small) $\varsigma_f^{(a)}$ factors so that $y_f^{(a)} = \varsigma_f^{(a)} m_f$ have acceptable magnitudes in the perturbative regime.

In the fermion weak-eigenstate basis, the relation between the Yukawa matrices $Y_f^{\prime (a)}$ and $M_f'$ involves the alignment factors
\beq
\varsigma^{\prime (a)}_{f}\, =\, U^{f}_L\, \varsigma_{f}^{(a)}\, U^{f\dagger}_L\, ,
\eeq
which, in general, are no-longer diagonal. Therefore, $Y_f^{\prime (a)}$ and $M'_f$ do not necessarily commute. The absence of FCNC interactions only requires this commutator to be zero in the fermion mass-eigenstate basis.

\section{Renormalization group equations}
\label{sec:rge}

The renormalization flow of the Yukawa couplings in a generic two-Higgs-doublet model was studied in Refs.~\cite{Cvetic:1997zd,Cvetic:1998uw}. The extension to a multi-Higgs-doublet model was first analysed in the lepton sector, neglecting all quark contributions ($\Gamma_a = \Delta_a = 0$) \cite{Grimus:2004yh}, and later extended to the most general case in Ref.~\cite{Ferreira:2010xe}. At the one-loop level, the Yukawa structures in Eq.~\eqn{eq:Yukawa} satisfy the RGEs \cite{Ferreira:2010xe,Bijnens:2011gd}:
\beqn
\cD \Gamma_a &=& a_\Gamma\, \Gamma_a
\, +\, \sum_{b=1}^N \left[
N_C\; \Tr \left( \Gamma_a \Gamma_b^\dagger
+ \Delta_a^\dagger \Delta_b \right)
+ \Tr \left(\Pi_a \Pi_b^\dagger \right) \right] \Gamma_b
\no \\ & + &
\sum_{b=1}^N \left(
- 2\, \Delta_b \Delta_a^\dagger \Gamma_b
+ \Gamma_a \Gamma_b^\dagger \Gamma_b
+ \frac{1}{2}\, \Delta_b \Delta_b^\dagger \Gamma_a
+ \frac{1}{2}\, \Gamma_b \Gamma_b^\dagger \Gamma_a \right) ,
\label{eq:Gamma}
\\
\cD \Delta_a &=& a_\Delta\, \Delta_a
\, +\, \sum_{b=1}^N \left[
N_C\; \Tr \left( \Delta_a \Delta_b^\dagger
+ \Gamma_a^\dagger \Gamma_b \right)
+ \Tr \left(\Pi_a^\dagger \Pi_b \right) \right] \Delta_b
\no\\ & + &
\sum_{b=1}^N \left(
- 2\, \Gamma_b \Gamma_a^\dagger \Delta_b
+ \Delta_a \Delta_b^\dagger \Delta_b
+ \frac{1}{2}\, \Gamma_b \Gamma_b^\dagger \Delta_a
+ \frac{1}{2}\, \Delta_b \Delta_b^\dagger \Delta_a \right) ,
\label{eq:Delta}
\\
\cD \Pi_a &=& a_\Pi\, \Pi_a
\, +\, \sum_{b=1}^N \left[
N_C\; \Tr \left( \Gamma_a \Gamma_b^\dagger
+ \Delta_a^\dagger \Delta_b \right)
+ \Tr \left(\Pi_a \Pi_b^\dagger \right) \right] \Pi_b
\no\\ & + &
\sum_{b=1}^N \left( \Pi_a \Pi_b^\dagger \Pi_b
+ \frac{1}{2}\, \Pi_b \Pi_b^\dagger \Pi_a \right) ,
\label{eq:Pi}
\eeqn
where $\cD\equiv 16\pi^2\mu\, (d/d\mu)$, being $\mu$ the renormalization scale, and $N_C=3$ is the number of quark colours. 

The gauge-boson corrections are incorporated through the factors
\beq
\label{a}
a_\Gamma \, =\, - 8\, g_s^2 - \frac{9}{4}\, g^2 - \frac{5}{12}\, {g^\prime}^2\, , 
\qquad\quad
a_\Delta \, =\, a_\Gamma - {g^\prime}^2 \, , 
\qquad\quad
a_\Pi \, =\, - \frac{9}{4}\, g^2 - \frac{15}{4}\, {g^\prime}^2\, ,
\eeq
where $g_s$, $g$ and $g'$ are the $SU(3)_{C}$, $SU(2)_L$ and $U(1)_Y$ couplings, respectively. These contributions do not change the flavour structure and only amount to a multiplicative global factor.

\begin{figure}[t]
\centerline{\mbox{}\hskip -.5cm
\begin{minipage}[b]{4.3cm}
\centering
\begin{tikzpicture}[line width=1.2 pt, node distance=1 cm and 1.5 cm] 
\coordinate[label= above: $\phi_b$] (phii); 
\coordinate[right = 1cm of phii] (aux);
\coordinate[right = of aux] (aux2);
\coordinate[right = 1cm of aux2, label = above: $\phi_a$] (phij); 
\draw[scalarnoarrow] (phii)--(aux);
\semiloop[fermion]{aux}{aux2}{0};
\semiloop[fermion]{aux2}{aux}{180};
\draw[scalarnoarrow](aux2)--(phij);
\draw[fill=black] (aux) circle (.07cm);
\draw[fill=black] (aux2) circle (.07cm);
\end{tikzpicture}
\vskip .32cm (a)
\end{minipage}
\hskip 1.1cm
\begin{minipage}[b]{5.2cm}
\centering
\begin{tikzpicture}[line width=1.2 pt, node distance=1 cm and 1.5 cm] 
\coordinate[label= above:${Q_{L}, q_{R}}\; $] (ua); 
\coordinate[right = 1cm of ua] (aux);
\coordinate[right = of aux] (aux2);
\coordinate[right = 1cm of aux2, label = above: $\; {Q_{L}, q_R}$] (ub); 
\coordinate[above = 0.7cm of aux2, label = above left: $\phi_b$](aux4);
\coordinate[right = 0.05cm of aux, label = below right : ${q_R , Q_L}$](aux5);
\draw[fermion] (ua)--(aux);
\draw[fermion] (aux)--(aux2);
\draw[fermion] (aux2)--(ub);
\semiloop[scalarnoarrow]{aux}{aux2}{0};
\draw[fill=black] (aux) circle (.07cm);
\draw[fill=black] (aux2) circle (.07cm);
\end{tikzpicture}
\vskip .42cm (b)
\end{minipage}
\hskip 1.1cm
\begin{minipage}[b]{4.5cm}
\centering
\begin{tikzpicture}[line width=1.2 pt, node distance=.9 cm and 1.5 cm, scale=0.9,transform shape]
\coordinate[label= above: $\phi_a$] (phii); 
\coordinate[right = 1cm of phii] (aux);
\coordinate[above right  = of aux](aux2);
\coordinate[below right  = of aux](aux3);
\coordinate[right  = of aux, label= right:$\phi_b$](aux4);
\coordinate[right = 1cm of aux2, label = right: $Q_L$ ](ub);
\coordinate[right = 1cm of aux3, label = right: ${u_R, d_R}$](ua);
\draw[scalarnoarrow] (phii)--(aux);
\draw[fermion] (aux)--(aux2);
\draw[fermionbar] (aux)--(aux3);
\draw[scalarnoarrow] (aux2)--(aux3);
\draw[fermion] (aux2)--(ub);
\draw[fermionbar] (aux3)--(ua);
\draw[fill=black] (aux) circle (.07cm);
\draw[fill=black] (aux2) circle (.07cm);
\draw[fill=black] (aux3) circle (.07cm);
\end{tikzpicture}
(c)
\end{minipage}}
\caption{One-loop topologies generating the flavour structures in Eqs.~\eqn{eq:Gamma}, \eqn{eq:Delta} and \eqn{eq:Pi}: scalar self-energies (a), $Q_L$ and $q_R$ self energies (b), and vertex corrections (c).}
\label{fig:RenYw}
\end{figure}
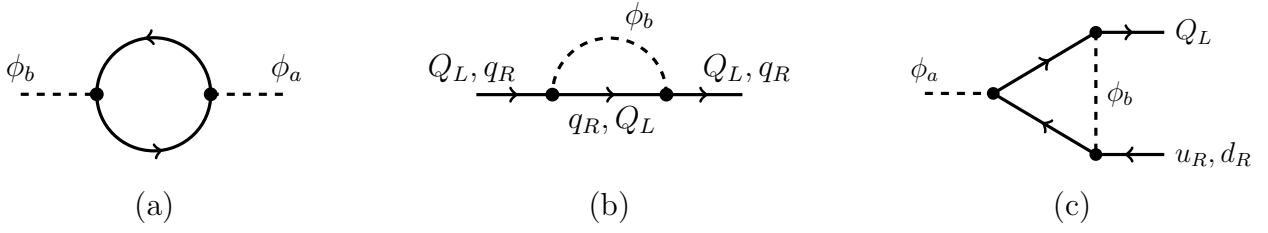

One-loop diagrams involving scalar propagators introduce two additional Yukawa matrices. The terms where these two matrices are traced
(first lines in the right-hand sides of Eqs.~\eqn{eq:Gamma}, \eqn{eq:Delta} and \eqn{eq:Pi})
originate in the scalar self-energies (Fig.~\ref{fig:RenYw}a). They correct each Yukawa vertex $\Gamma_b$, $\Delta_b$, $\Pi_b$ with a different multiplicative factor, 
leaving untouched its own flavour configuration, and mix the different `$b$' structures. The additional flavour-dependent quantum corrections in the second lines arise from fermion self-energies and vertex contributions. The $Q_L$ self-energy (Fig.~\ref{fig:RenYw}b) generates the $(\Gamma_b \Gamma_b^\dagger + \Delta_b \Delta_b^\dagger)$
terms multiplying the left-hand sides of $\Gamma_a$ in \eqn{eq:Gamma} and $\Delta_a$ in \eqn{eq:Delta}, while the $d_R$ and $u_R$ self-energies (Fig.~\ref{fig:RenYw}b) give rise to the $\Gamma_a \Gamma_b^\dagger \Gamma_b$  and $\Delta_a \Delta_b^\dagger \Delta_b$ contributions, respectively. The
vertex topology  (Fig.~\ref{fig:RenYw}c) introduces the remaining structures $\Delta_b \Delta_a^\dagger \Gamma_b$ and $\Gamma_b \Gamma_a^\dagger \Delta_b$, with `$b$' indices in both sides of the primary `$a$' Yukawa. The corresponding terms in $\cD\Pi_a$ are easily obtained with the changes $\Gamma_a\to\Pi_a$, $\Delta_a\to 0$. We have recalculated all these topologies, finding complete agreement with Refs.~\cite{Ferreira:2010xe,Bijnens:2011gd}. 

Let us now consider a tree-level Yukawa structure having the generalized aligned-like form of Eq.~\eqn{eq:alignment} with $\varsigma_f^{(a)}$ diagonal matrices. Focusing for the moment on the $\Gamma_a$ couplings, one can rewrite Eq.~\eqn{eq:Gamma} as
\beq\label{eq:rgeStruc}
\cD\Gamma_a\, =\, \e^{-i\tilde\theta_a}\,\left\{ \xi_d^{(a)}\; \cD\Gamma_1 + 
 \left[\delta\xi_d^{(a)}
  + \Theta_{d,FC}^{(a)} + \Theta_{d,FV}^{(a)}\right]\, \Gamma_1\right\}\, .
\eeq
The parameters $\delta\xi_d^{(a)}$ contain those terms in the first line of Eq.~\eqn{eq:Gamma} which do not fit in $\xi_d^{(a)} \cD\Gamma_1$. Since they are constants without flavour structure, these contributions can be reabsorbed into a quantum redefinition of the alignment factors, $\e^{-i\tilde\theta_a}\,\delta\xi_d^{(a)} = \cD \left(\e^{-i\tilde\theta_a}\,\xi_d^{(a)}\right)$, promoting them to $\mu$-dependent quantities. The contributions from the second line of Eq.~\eqn{eq:Gamma} have been split in two parts:
$\Theta_{d,FC}^{(a)}$ incorporates the flavour-conserving terms with $\Gamma_b$ structures, while $\Theta_{d,FV}^{(a)}$ contains the flavour-violating pieces with $\Delta_b$ matrices.

A similar decomposition can be performed for $\cD\Delta_a$ and $\cD\Pi_a$. Obviously, one does not generate any FCNC couplings through $\cD\Pi_a$ because there is only one flavour structure in the second line of \eqn{eq:Pi} (in aligned-like models), {\it i.e.},
$\Theta_{\ell,FV}^{(a)}=0$.

Since we are only interested in the flavour-violating structures, we can neglect the quantum corrections to the vacuum expectation values and work directly in the Higgs basis where all expressions simplify considerably. Dropping all flavour-conserving contributions, the integration of the RGEs is quite straightforward.
At leading order, one gets the following local FCNC interactions 
(in the neutral scalar mass eigenstates basis):
\beqn\label{eq:FCNC}
\cL_{\mathrm{FCNC}} &\!\! =&\!\! \frac{1}{4\pi^2v^3}\,\sum_{k=1}^{2N-1}  \varphi_k^0
\; \sum_{a=1}^{N-1} \left\{ \cC_d^{(a+1)}\,  \left( \cR_{k, 2a}\, + i \,\cR_{k, 2a+1}\right)\, \bar d_L \widetilde\Theta_{d}^{(a+1)} M_d^{\phantom{\dagger}} d_R\, \right.
\no \\ &  & \left. \mbox{} \hskip 3.2cm
+\:  \cC_u^{(a+1)}\, \left( \cR_{k, 2a}\, - i \,\cR_{k, 2a+1}\right)\,\bar u_L \widetilde\Theta_{u}^{(a+1)} M_u^{\phantom{\dagger}} u_R
\right\} +\;\mathrm{h.c.}\, ,\quad
\eeqn
where each quark vertex is proportional to the corresponding mass.
The structures
\begin{align}\label{eq:Theta}
\widetilde\Theta_{d}^{(a)} &= \;
- V^\dagger_{_\mathrm{CKM}} \sum_{b=1}^N 
\varsigma^{(b)\dagger}_u M_u^{\phantom{\dagger}} M_u^\dagger\varsigma^{(a)}_u V_{_\mathrm{CKM}}^{\phantom{\dagger}}\varsigma_d^{(b)}
+ \varsigma_d^{(a)}V^\dagger_{_\mathrm{CKM}} \sum_{b=1}^N
\varsigma^{(b)\dagger}_u M_u^{\phantom{\dagger}} M_u^\dagger V_{_\mathrm{CKM}}^{\phantom{\dagger}} \varsigma^{(b)}_d
+ \Delta \widetilde\Theta_{d}^{(a)}\, ,
\no\\ \\
\widetilde\Theta_{u}^{(a)} &= \; 
- V_{_\mathrm{CKM}}^{\phantom{\dagger}} \sum_{b=1}^N 
\varsigma^{(b)}_d M_d^{\phantom{\dagger}} M_d^\dagger\varsigma^{(a)\dagger}_d V^\dagger_{_\mathrm{CKM}}\varsigma_u^{(b)\dagger}
+ \varsigma_u^{(a)\dagger}V_{_\mathrm{CKM}}^{\phantom{\dagger}} \sum_{b=1}^N
\varsigma^{(b)}_d M_d^{\phantom{\dagger}} M_d^\dagger V^\dagger_{_\mathrm{CKM}} \varsigma^{(b)\dagger}_u
+ \Delta \widetilde\Theta_{u}^{(a)}\, ,
\no
\end{align}
involve two additional quark mass matrices, two CKM mixing matrices and three alignment factors. Thus, the generated FCNC operators have dimension seven and are strongly suppressed by CKM mixings. The last terms in \eqn{eq:Theta},
\begin{align}
\label{eq:DTheta}
\Delta\widetilde\Theta_{d}^{(a)} &=\; \frac{1}{4}\, \left[
V^\dagger_{_\mathrm{CKM}}\left(\sum_{b=1}^N \varsigma^{(b)\dagger}_u M_u^{\phantom{\dagger}} M_u^\dagger \varsigma^{(b)}_u\right) V_{_\mathrm{CKM}}^{\phantom{\dagger}}\, ,\, \varsigma_d^{(a)}\right]
& =\; \frac{N}{4}\, \left[ V^\dagger_{_\mathrm{CKM}} M_u^{\phantom{\dagger}} M_u^\dagger
V_{_\mathrm{CKM}}^{\phantom{\dagger}}\, ,\, \varsigma_d^{(a)}\right]\, ,
\no \\ \\ 
\Delta\widetilde\Theta_{u}^{(a)} & =\; \frac{1}{4}\, \left[
V_{_\mathrm{CKM}}^{\phantom{\dagger}}\left(\sum_{b=1}^N \varsigma^{(b)}_d M_d^{\phantom{\dagger}} M_d^\dagger\varsigma^{(b)\dagger}_d\right) V^\dagger_{_\mathrm{CKM}}\, ,\, \varsigma_u^{(a)\dagger}\right]
&=\; \frac{N}{4}\, \left[  V_{_\mathrm{CKM}}^{\phantom{\dagger}} M_d^{\phantom{\dagger}} M_d^\dagger V_{_\mathrm{CKM}}^{\phantom{\dagger}}\, ,\, \varsigma_d^{(a)}\right] \, ,
\no
\end{align}
are only present in the most general aligned-like scenario with diagonal matrices $\varsigma_f^{(a)}$, otherwise the commutators would vanish identically.

In the simpler case of normal alignment, where the factors $\varsigma_f^{(a)}$ are just family-universal parameters, these expressions adopt the much simpler forms:
\beqn\label{eq:Theta_d_alig}
\widetilde\Theta_{d}^{(a)} & = & \left( \varsigma_d^{(a)} - \varsigma^{(a)}_u\right)\,\left( \sum_{b=1}^N \varsigma^{(b)\dagger}_u\varsigma^{(b)}_d\right)\;
V^\dagger_{_\mathrm{CKM}} M_u^{\phantom{\dagger}} M_u^\dagger V_{_\mathrm{CKM}}^{\phantom{\dagger}}\, ,
\\ \label{eq:Theta_u_alig}
\widetilde\Theta_{u}^{(a)} & = &
\left(\varsigma_u^{(a)\dagger} - \varsigma^{(a)\dagger}_d \right)\,\left(\sum_{b=1}^N \varsigma^{(b)\dagger}_u\varsigma^{(b)}_d\right)\;
V_{_\mathrm{CKM}}^{\phantom{\dagger}} M_d^{\phantom{\dagger}} M_d^\dagger V^\dagger_{_\mathrm{CKM}}\, .
\eeqn
For $N=2$, these results agree with the previously known one-loop misalignment of the A2HDM \cite{Ferreira:2010xe,Jung:2010ik,Botella:2015yfa,Braeuninger:2010td,Bijnens:2011gd,Li:2014fea,Abbas:2015cua,Gori:2017qwg}.

The RGEs determine the $\mu$ dependence of the Wilson coefficients $\cC_{d,u}^{(a)}(\mu)$. At leading order, one finds ($f=d,u$)
\beq \label{eq:mudep}
\cC_f^{(a)}(\mu)\, =\, \cC_f^{(a)}(\mu_0) - \log{(\mu/\mu_0)}\, .
\eeq

One can easily check that $\cL_{\mathrm{FCNC}}$ vanishes identically for all models with natural flavour conservation, discussed in section~\ref{sec:NFC}. Each of these models is characterized by three numbers $\{a_d,a_u,a_\ell\}$, specifying the choice of three scalar fields coupling to the different types of right-handed fermions, and real alignment parameters $\varsigma_f^{(a)} = \Omega_{aa_f}/\Omega_{1a_f}$. Therefore,
\beq
\left( \varsigma_d^{(a)} - \varsigma^{(a)}_u\right)\, \sum_{b=1}^N \varsigma^{(b)}_u\varsigma^{(b)}_d\, =\, 
\left( \Omega_{a a_d} - \Omega_{a a_u}\right)\, 
\frac{\sum_{b=1}^N \Omega_{ba_u}\Omega_{ba_d}}{\left(\Omega_{1a_u}\Omega_{1a_d}\right)^2}\, =\, 
\left( \Omega_{a a_d} - \Omega_{a a_u}\right)\, 
\frac{\delta_{a_u a_d}}{\left(\Omega_{1a_u}\Omega_{1a_d}\right)^2}\, =\,
0\, ,
\eeq
which implies $\widetilde\Theta_{d}^{(a)} = \widetilde\Theta_{u}^{(a)} = 0$.

The one-loop FCNC local interactions also disappear if the Yukawa matrices satisfy the relations
\beq\label{eq:QuasiAlignment}
\bat\displaystyle
\sum_{b=1}^N\Delta_b\Delta_a^\dagger\Gamma_b\, =\, \lambda_\Gamma\, \Gamma_a\, ,
\qquad\qquad\qquad &\displaystyle
\sum_{b=1}^N\Delta_b\Delta_b^\dagger\Gamma_a\, =\, \lambda'_\Gamma\, \Gamma_a\, ,
\\[20pt]\displaystyle
\sum_{b=1}^N\Gamma_b\Gamma_a^\dagger\Delta_b\, =\, \lambda_\Delta\, \Delta_a\, ,
\qquad\qquad\qquad &\displaystyle
\sum_{b=1}^N\Gamma_b\Gamma_b^\dagger\Delta_a\, =\, \lambda'_\Delta\, \Delta_a\, ,
\ea
\eeq
with $\lambda_\Gamma$, $\lambda'_\Gamma$, $\lambda_\Delta$, $\lambda'_\Delta$
arbitrary complex parameters. In this very particular case, $\cL_{\mathrm{FCNC}}$ becomes flavour conserving. The conditions \eqn{eq:QuasiAlignment} have been analysed in Ref.~\cite{Botella:2015yfa}, within the A2HDM, finding a phenomenologically viable solution with all Yukawa matrices proportional to the ``democratic'' matrix $\mathcal{Y}_{ij} = 1\, , \forall i,j$. This stable aligned solution is protected by a $\cZ_3\otimes \cZ'_3$ symmetry and corresponds to the limit where only one generation of quarks (top and bottom) acquires mass, while  
$V_{_\mathrm{CKM}}$ is the identity matrix.

\section{Flavour symmetries}
\label{sec:symmetry}

The flavour structure of $\cL_{\mathrm{FCNC}}$ can be easily understood with symmetry considerations \cite{Pich:2009sp}. In the absence of Yukawa couplings, the Lagrangian of the N-Higgs-doublet model has a huge $SU(3)^5$ flavour symmetry, corresponding to independent transformations of the $Q_L$, $L_L$, $d_R$, $u_R$ and $\ell_R$ fermion fields in the 3-generation flavour space: $f_X\to S_{f_X}\, f_X\, , \; S_{f_X}\in SU(3)_{f_X}$. One can formally extend this symmetry to the Yukawa sector, assigning appropriate transformation properties to the flavour matrices $\Gamma_a$, $\Delta_a$ and $\Pi_a$, which are then treated as spurion fields \cite{Chivukula:1987py,DAmbrosio:2002vsn}:
\beq\label{eq:MFV}
\Gamma_a\;\to\; S_{Q_L}\, \Gamma_a\, S_{d_R}^\dagger\, ,\qquad 
\Delta_a\;\to\; S_{Q_L}\, \Delta_a\, S_{u_R}^\dagger\, ,\qquad 
\Pi_a\;\to\; S_{L_L}\, \Pi_a\, S_{\ell_R}^\dagger\, . 
\eeq
These auxiliary fictitious fields allow for an easy bookkeeping of operators invariant under the enlarged symmetry, and encode the explicit symmetry breakings introduced by the Yukawa interactions. Obviously, the renormalization group equations~\eqn{eq:Gamma}, \eqn{eq:Delta} and \eqn{eq:Pi} transform homogeneously under \eqn{eq:MFV} because quantum corrections respect the Lagrangian symmetries (modulo anomalies). Only those structures which are invariant under this formal flavour symmetry can be generated at higher orders.

Once the symmetry breakings are explicitly included, the Yukawa Lagrangian~\eqn{eq:YL} remains still invariant under flavour-dependent phase transformations of the fermion mass eigenstates, provided one performs appropriate
rephasings of all flavour structures (masses, Yukawa couplings and quark-mixing factors)
\cite{Pich:2009sp,Pich:2010ic,Jung:2010ik}:
\beq\begin{array}{ccc}
f^i_X\;\to\; \e^{i \alpha_i^{f,X}}\; f^i_X\, ,
&\qquad\qquad &
Y_f^{(a),ij}\;\to\; \e^{i \alpha_i^{f,L}}\; Y_f^{(a),ij}\; \e^{-i \alpha_j^{f,R}}\, ,
\\[7pt]
M_f^{ij}\;\to\; \e^{i \alpha_i^{f,L}}\;
M_f^{ij}\; \e^{-i \alpha_j^{f,R}}\, ,
&&
V_{_\mathrm{CKM}}^{ij}\;\to\; \e^{i \alpha_i^{u,L}}\;
V_{_\mathrm{CKM}}^{ij}\; \e^{-i \alpha_j^{d,L}}\, .
\ea
\eeq
Here, $f=d,u,\ell$, $X=L,R$ and $i,j$ refer to the three different fermion families. The generalized alignment condition \eqn{eq:GAlignment} implies then
\beq
\varsigma_f^{(a),ij}\;\to\; \e^{i \alpha_i^{f,L}}\;
\varsigma_f^{(a),ij}\; \e^{-i \alpha_j^{f,L}}\, .
\eeq

Since quantum corrections preserve these flavour symmetries, they can only give rise to
FCNC operators of the form
\begin{align}\label{eq:operators}
\cO_d^{n,m} & = \; \bar d_L (\varsigma_d^{\phantom{\dagger}})^{p_1} V_{_\mathrm{CKM}}^\dagger
(\varsigma_u^\dagger)^{p_n} (M_u^{\phantom{\dagger}} M_u^\dagger)^n (\varsigma_u^{\phantom{\dagger}})^{p'_{n}}
V_{_\mathrm{CKM}}^{\phantom{\dagger}} (\varsigma_d^{\phantom{\dagger}})^{p_m} (M_d^{\phantom{\dagger}} M_d^\dagger)^m (\varsigma_d^\dagger)^{p'_{m}}
(\varsigma_d^{\phantom{\dagger}})^{p'_1} M_d^{\phantom{\dagger}} d_R\, ,\quad
\no\\ \\
\cO_u^{n,m} & = \; \bar u_L (\varsigma_u^{\phantom{\dagger}})^{p_1 }
V_{_\mathrm{CKM}}^{\phantom{\dagger}} (\varsigma_d^{\phantom{\dagger}})^{p_n} (M_d^{\phantom{\dagger}} M_d^\dagger)^n (\varsigma_d^\dagger)^{p'_{n}}
V_{_\mathrm{CKM}}^\dagger
(\varsigma_u^\dagger)^{p_m} (M_u^{\phantom{\dagger}} M_u^\dagger)^m (\varsigma_u^{\phantom{\dagger}})^{p'_{m}} (\varsigma_u^\dagger)^{p'_1}
M_u^{\phantom{\dagger}} u_R\, ,
\no 
\end{align}
or similar structures with additional factors of $V_{_\mathrm{CKM}}^{\phantom{\dagger}}$,  
$V_{_\mathrm{CKM}}^\dagger$, $(M_f^{\phantom{\dagger}} M_f^\dagger)$ and alignment matrices. To generate a FCNC operator one needs at least two insertions of the CKM mixing matrix, and the unitarity of $V_{_\mathrm{CKM}}^{\phantom{\dagger}}$ requires the presence of quark mass matrices between these two insertions, {\it i.e.}, a product $(M_f^{\phantom{\dagger}} M_f^\dagger)^n$ with $n\ge 1$. An additional (single) mass factor is needed at the end of the chain to preserve chirality. Thus,  the lowest-order operators must contain two quark-mixing matrices and three mass matrices, as explicitly shown in Eq.~\eqn{eq:FCNC}.

The alignment factors originate in the Yukawa matrices $Y^{(a)}_f = \varsigma^{(a)}_f M_f$. Since $\varsigma^{(1)}_f =1$, the terms $(\varsigma_f^{\phantom{\dagger}})^{p_k,p'_k}$ and $(\varsigma_f^{\dagger})^{p_k,p'_k}$ in \eqn{eq:operators} refer to the possible presence of $p_k , p_{k'}\le k$ non-trivial alignment parameters with possibly different values of the superindex $(a)$. To simplify notation, we have loosely skipped this superindex and have made use
of the commutation property of the diagonal matrices $M_f^{\phantom{\dagger}}$ and
$ \varsigma^{(a)}_f$ (in the fermion-mass eigenstate basis) to collect together alignment factors of a given type.
Thus, the operators $\widetilde\Theta_{d}^{(a)}$ and $\widetilde\Theta_{u}^{(a)}$ in Eq.~\eqn{eq:Theta} contain up to three alignment factors. Notice that alignment structures with $b\not = a$ can only appear pairwise, $\varsigma_f^{(b)} \varsigma_{f'}^{(b)\dagger}$, since they are generated through the exchange of a scalar propagator between two `$b$' Yukawa vertices. 

The first possible alignment factor in the r.h.s of Eqs.~\eqn{eq:operators}, just before the first CKM matrix, has a more subtle origin. It compensates the   $\varsigma_d^{(a)} {\cal D}\Gamma_1$ terms in Eq.~\eqn{eq:rgeStruc} which are not present in ${\cal D}\Gamma_2$, and the $\varsigma_u^{(a)\dagger} {\cal D}\Delta_1$ terms not present in ${\cal D}\Delta_2$. Therefore, in this position there is at most a single alignment factor which must be either  $\varsigma_d^{(a)}$ or $\varsigma_u^{(a)\dagger}$, for $\cO_d^{n,m}$ and $\cO_u^{n,m}$, respectively, as explicitly shown in Eqs.~\eqn{eq:Theta}.

\section{Phenomenological constraints}
\label{sec:phenomenology}

In the absence of protecting $\cZ_2$ symmetries, the alignment hypothesis can only be exactly fulfilled at a single value of the renormalization scale $\mu=\Lambda_A$. Quantum corrections unavoidably misalign the Yukawa matrices at $\mu\not= \Lambda_A$, generating FCNC vertices that contribute to processes which are very suppressed in the SM. However, the flavour symmetries embodied in the tree-level aligned Lagrangian restrict very efficiently the possible structures that can be generated at higher perturbative orders. At the one-loop level, the resulting FCNC local interaction in Eq.~\eqn{eq:FCNC} only contains two operators, one for each quark sector, up or down. Both operators contain two insertions of the CKM matrix and three Yukawa matrices, which entails a strong phenomenological suppression of FCNC effects. Nevertheless, it is worth to investigate whether any interesting contributions could still show up at a level relevant for present or forthcoming experiments.

For simplicity, from now on we will restrict the analysis to the usual A2HDM framework, {\it i.e.}, a two-Higgs-doublet Lagrangian with aligned Yukawa structures, parametrized with three alignment constants $\varsigma_{d,u,\ell}$.
The one-loop FCNC effective Lagrangian~\eqn{eq:FCNC} reduces in this case to \cite{Jung:2010ik}
\beqn\label{eq:FCNC2}
\cL_{\mathrm{FCNC}} &\!\!\! =&\!\!\! \frac{1}{4\pi^2v^3}\,
\left( 1 + \varsigma_u^* \varsigma_d^{\phantom{*}}\right)\,
\sum_{k=1}^{3}  \varphi_k^0
\, \left\{ \cC_d(\mu)\,  \left( \cR_{k2}\, + i \,\cR_{k3}\right) 
\left( \varsigma_d^{\phantom{*}} - \varsigma_u^{\phantom{*}} \right)\,
\bar d_L V^\dagger_{_\mathrm{CKM}} M_u^{\phantom{\dagger}} M_u^\dagger V_{_\mathrm{CKM}}^{\phantom{\dagger}} M_d^{\phantom{\dagger}} d_R\, \right.
\no \\ &  & \left. \mbox{} \hskip 1.cm
-\:  \cC_u(\mu)\, \left( \cR_{k2}\, - i \,\cR_{k3}\right)
\left( \varsigma_d^* - \varsigma_u^* \right)\,
\bar u_L V_{_\mathrm{CKM}}^{\phantom{\dagger}} M_d^{\phantom{\dagger}} M_d^\dagger V^\dagger_{_\mathrm{CKM}} M_u^{\phantom{\dagger}} u_R
\right\} 
\; +\;\mathrm{h.c.}
\eeqn
with $\cC_{d,u}(\mu)$ encoding the renormalization-scale dependence, which at leading order takes the simple form: $\cC_{d,u}(\mu) = \cC_{d,u}(\mu_0) - \log{(\mu/\mu_0)}$.

The sum runs over the three neutral scalars of the model. Assuming that CP is a symmetry of the scalar potential (and vacuum), there are two CP-even neutral scalars ($\varphi^0_1 = h, \, \varphi^0_2 = H$) which mix through a two-dimensional rotation matrix, while the third neutral scalar $\varphi_3^0 =A$ is CP-odd and does not mix with the others. Therefore:
\beq\label{eq:RmixingCP}
\cR_{11} =\cR_{22} =\cos{\tilde\alpha}\, ,\quad
\cR_{12} = - \cR_{21} =\sin{\tilde\alpha}\, ,\quad
\cR_{33} = 1\, ,\quad
\cR_{13} =\cR_{23} =\cR_{31} =\cR_{32} = 0\, .
\eeq
We adopt the convention $0\le \tilde\alpha\le\pi$, so that $\sin{\tilde\alpha}$ is always positive, and will identify the CP-even neutral state $h$ with the Higgs particle found at LHC, {\it i.e.}, $M_h = (125.09 \pm 0.24)\:\mathrm{GeV}$ \cite{Aad:2015zhl}. The data shows that $h$ behaves like the SM Higgs boson, within the current experimental uncertainties, which constraints the mixing angle to satisfy $|\cos{\tilde\alpha}| > 0.90$  (68\% CL) \cite{Celis:2013rcs,Celis:2013ixa}.

One could speculate that flavour alignment originates in some underlying new-physics dynamics at a high-energy scale $\Lambda_A$, where alignment is exact due to a flavour symmetry of the new-physics Lagrangian, {\it i.e.}, $\cC_f(\Lambda_A)=0$. Several models with this property have been discussed in the literature \cite{Serodio:2011hg,Varzielas:2011jr,Cree:2011uy,Celis:2014zaa,Knapen:2015hia}. In that case, the RGEs determine $\cC_f(\mu) = \log{(\Lambda_A/\mu)}$ at an arbitrary renormalization scale $\mu$. Taking $\Lambda_A \le M_{\mathrm{Planck}} \sim 10^{19}\:\mathrm{GeV}$, one gets\ $\cC_f(M_W)\le 40$, which puts an upper bound on the size of any possible FCNC effects. Tree-level implications of $\cL_{\mathrm{FCNC}}$ have been already analysed in Refs.~\cite{Braeuninger:2010td,Gori:2017qwg}, with the extreme choice $\Lambda_A = M_{\mathrm{Planck}}$, while different values of the high-energy scale $\Lambda_A$ were investigated in Ref.~\cite{Bijnens:2011gd}. 

While being illustrative of the possible phenomenological relevance of the Yukawa misalignment, the simplified tree-level analyses completely neglect the non-local FCNC loop contributions generated by the A2HDM Lagrangian \cite{Jung:2010ik,Li:2014fea,Abbas:2015cua,Jung:2010ab,Jung:2012vu,Chang:2015rva,Hu:2016gpe,Hu:2017qxj,Cho:2017jym,Enomoto:2015wbn}, which are usually dominant. The most important FCNC processes originate in one-loop diagrams (penguins and boxes) involving charged-current flavour-changing vertices, through the exchange of $W^\pm$ gauge bosons and the unique charged scalar ($\varphi_1^\pm = H^\pm$) present in the model. Most of these loop contributions generate finite amplitudes (also at higher orders) because symmetry considerations forbid the presence of the relevant FCNC counterterms in the Lagrangian. This is no-longer true for the effective FCNC interactions of the neutral scalars; the loop contributions generate in this case ultraviolet (UV) divergences that get exactly cancelled through the renormalization of the $\cC_f$ couplings in Eq.~\eqn{eq:FCNC2} (and similar counterterms at higher orders). The renormalization-scale dependence of the loop contributions cancels also the $\mu$ dependence of the $\cC_f(\mu)$ misalignment parameters. Complete one-loop calculations, including the proper renormalization of the misalignment Lagrangian $\cL_{\mathrm{FCNC}}$ have been already published for the FCNC transitions $B^0_{d,s} \rightarrow \ell^+ \ell^-$ \cite{Li:2014fea} and $t\to \varphi^0_k c$ \cite{Abbas:2015cua}.

Owing to the quark-mass and CKM suppressions of $\cL_{\mathrm{FCNC}}$ the potentially largest misalignment effects should appear in the $\varphi_k^0\bar s_L b_R$ effective vertex, with a top contribution proportional to $V_{ts}^* V_{tb}^{\phantom{*}} m_t^2 m_b/(4\pi^2 v^3)$. In the absence of any direct evidence of FCNC Higgs decays, this singles out $B_{s}^0 \rightarrow \mu^+ \mu^-$ and $B_{s}^0$--$\bar{B}_s^0$ mixing as prime candidates to test the local FCNC interaction. As shown in Fig.~\ref{fig:diagrams}, both processes get tree-level contributions from $\cL_{\mathrm{FCNC}}$, through $\varphi_k^0$ exchange. There is, however, an important difference between the two transitions. The leptonic $B_s^0\rightarrow \mu^+ \mu^-$ decay occurs with a single insertion of the effective $\varphi_k^0\bar s_L b_R$ vertex which, therefore, renormalizes the corresponding one-loop scalar-penguin contribution \cite{Li:2014fea}. On the other side, to generate a $B_{s}^0$--$\bar{B}_s^0$ mixing transition through neutral scalar exchange, one needs to insert two FCNC effective vertices. This contribution is then of a higher-perturbative order and should be considered together with the relevant two-loop contributions to the meson-mixing amplitude, since it renormalizes the UV divergence from diagrams with two (one-loop) scalar-penguin triangles. The one-loop diagrammatic calculation of the meson-antimeson transition is in fact UV convergent \cite{Jung:2010ik}.

\begin{figure}[t]
\centering
\begin{tikzpicture}[line width=1.2 pt,node distance=.9 cm and 1.2 cm]
\coordinate[] (v1);
\coordinate[above left = of v1,  label = left: $b$] (q1);
\coordinate[below left = of v1,  label = left: ${\bar{d}, \bar{s}}$] (q2);
\coordinate[right = 1.5 cm of v1] (v2);
\coordinate[below right = of v2, label = right: $\mu^+$] (l1);
\coordinate[above right = of v2, label = right: $\mu^-$] (l2);
\coordinate[right = 0.75cm of v1, label = above: $\varphi^0_k$] (phi);
\draw[scalarnoarrow] (v1) -- (v2);
\draw[fermion] (q1) -- (v1);
\draw[fermionbar] (q2) -- (v1);
\draw[fermion] (l1) -- (v2);
\draw[fermionbar] (l2) -- (v2);
\draw[fill=black] (v2) circle (.07cm);
\draw[fill=white] (v1) circle (.13cm);
\draw (v1) node[cross = .13cm] {};
\end{tikzpicture}
\hskip 2cm
\begin{tikzpicture}[line width=1.2 pt,node distance=.9 cm and 1.2 cm]
\coordinate[] (v1);
\coordinate[above left = of v1,  label = left: $b$] (q1);
\coordinate[below left = of v1,  label = left: ${\bar{d}, \bar{s}}$] (q2);
\coordinate[right = 1.5 cm of v1] (v2);
\coordinate[below right = of v2, label = right: $\bar{b}$] (l1);
\coordinate[above right = of v2, label = right: ${d, s}$] (l2);
\coordinate[right = 0.75cm of v1, label = above: $\varphi^0_k$] (phi);
\draw[scalarnoarrow] (v1) -- (v2);
\draw[fermion] (q1) -- (v1);
\draw[fermionbar] (q2) -- (v1);
\draw[fermion] (l1) -- (v2);
\draw[fermionbar] (l2) -- (v2);
\draw[fill=white] (v2) circle (.13cm);
\draw (v2) node[cross = .13cm] {};
\draw[fill=white] (v1) circle (.13cm);
\draw (v1) node[cross = .13cm] {};
\end{tikzpicture}
\caption{Feynman diagrams contributing to $\bar B^0 \rightarrow \mu^+ \mu^-$ (left) and  $B^0$--$\bar{B}^0$ mixing (right). The crossed vertex represents the one-loop effective FCNC neutral interaction in Eq.~\eqn{eq:FCNC}.}
\label{fig:diagrams}
\end{figure}
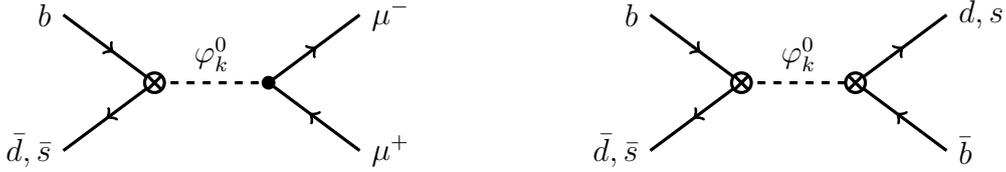

\subsection{Inputs and numerical treatment}
\label{sec:input}

We are interested in a scalar sector testable at the LHC, with the masses of the additional scalars not too far from the electroweak scale. 
A lower bound $M_{H^{\pm}}\ge 78.6\:\mathrm{GeV}$ (95\% CL) is imposed by LEP searches \cite{Searches:2001ac}, with the only assumption that the charged scalar decays into fermions. In addition, the precise measurements of the $Z$ and $W^\pm$ self-energies, usually encoded through the so-called oblique parameters $S$, $T$ and $U$ \cite{Peskin:1990zt}, impose strong constraints on the scalar mass splittings. Together with the requirement of perturbativity and perturbative unitary bounds on the scalar potential couplings \cite{Kanemura:2015ska}, this implies that the additional neutral scalars $H$ and $A$ should have masses below the TeV, if $ M_{H^{\pm}}<~500$~GeV~\cite{Celis:2013ixa}. 

In order to illustrate the possible phenomenological scenarios, we will adopt the following benchmark configurations for the unknown scalar masses:
\beq
\label{eq:massconf}
\begin{array}{llll}
\mathrm{A:}\quad & M_{H^{\pm}} = 100~\mathrm{GeV}\, , &  M_H = 50~\mathrm{GeV}\, , &  M_A = 50~\mathrm{GeV}\, ,
\\[2pt]
\mathrm{B:}\; & M_{H^{\pm}} = 100~\mathrm{GeV}\, , &  M_H = 200~\mathrm{GeV}\, , &  M_A = 200~\mathrm{GeV}\, ,
\\[2pt]
\mathrm{C:}\; & M_{H^{\pm}} = 500~\mathrm{GeV}\, , &  M_H = 500~\mathrm{GeV}\, , &  M_A = 200~\mathrm{GeV}\, ,
\\[2pt]
\mathrm{D:}\; & M_{H^{\pm}} = 500~\mathrm{GeV}\, , &  M_H = 200~\mathrm{GeV}\, , &  M_A = 500~\mathrm{GeV}\, ,
\\[2pt]
\mathrm{E:}\; & M_{H^{\pm}} = 1000~\mathrm{GeV}\, , &  M_H = 500~\mathrm{GeV}\, , &  M_A = 1000~\mathrm{GeV}\, ,
\\[2pt]
\mathrm{F:}\; & M_{H^{\pm}} = 1000~\mathrm{GeV}\, ,\quad &  M_H = 1000~\mathrm{GeV}\, ,\quad &  M_A = 1000~\mathrm{GeV}\, .
\ea
\eeq
These mass configurations satisfy the present experimental constraints on the oblique parameters \cite{Celis:2013ixa,Baak:2014ora}. The first four choices are representative of a plausible nearby scalar spectrum, while the last two approach the decoupling regime.

The up-type alignment parameter is strongly constrained by the measured $Z \rightarrow b \bar{b}$ decay width, which leads to an upper bound that scales  linearly with the charged scalar mass \cite{Jung:2010ik}:
\beq
\abs{\varsigma_u}\; <\; 0.72\, +\, 0.0024\; M_{H^{\pm}}/\mathrm{GeV}
\qquad (95\%\;\mathrm{CL})\, .
\eeq
With $ M_{H^{\pm}} \le 500$ GeV, this gives $\abs{\varsigma_u} < 1.9$ at 95\% CL.
For the other two alignment parameters we require the Yukawa couplings to remain in the perturbative regime, {\it i.e.}, $\frac{\sqrt{2}}{v} \varsigma_{f} m_f < 1$. This implies the absolute upper bounds $\abs{\varsigma_d} < 50$ and $\abs{\varsigma_\ell} < 100$. Our numerical analysis will be performed in the CP-conserving limit to reduce the number of free parameters.

The choice of CKM parameters is subtle because global CKM fits assume the SM. We have performed a specific fit to obtain the CKM elements needed for our analysis, taking as entries determinations which are not sensitive to new physics. First of all $V_{ud} $ is extracted from the $(0^+ \rightarrow 0^+)$ nuclear $\beta$ decays \cite{Hardy:2014qxa} and CKM unitarity is used to determine $V_{us}\equiv \lambda$. The value of $V_{ub}$ is obtained combining the exclusive and inclusive averages from $b \rightarrow u l \bar{\nu}_{l}$ decays, performed by the Heavy Flavor Averaging Group (HFLAV) \cite{Amhis:2016xyh}, and increasing the error with the usual PDG scale factor to account for their present discrepancy \cite{Patrignani:2016xqp}. For $V_{cb}$ we adopt the most recent inclusive fit to semileptonic $b \rightarrow c l \bar{\nu}_{l}$ data \cite{Gambino:2016jkc}, which turns out to be consistent with the latest exclusive determinations, once the uncertainties related with the adopted form-factor parametrizations are properly assessed \cite{Bigi:2016mdz,Bigi:2017njr,Bigi:2017jbd,Grinstein:2017nlq,Bernlochner:2017jka,Bernlochner:2017xyx}. Then, combining $V_{cb}$ with the previous value of $\lambda$, the Wolfenstein $A$ parameter is obtained.  The apex $(\bar{\rho}, \bar{\eta})$ of the `$bd$' unitarity triangle is determined from $V_{ub}/V_{cb}, \lambda$ and the ratio $\Delta m_{B_s^0}/\Delta m_{B_d^0}$, which fixes $V_{td}/V_{ts}$ \cite{Amhis:2016xyh}, by performing a $\chi^2$ minimization. These ratios are related to $\bar{\rho}$ and $\bar{\eta}$ through:
\begin{equation}
\abs{\frac{V_{ub}}{V_{cb}}}\, =\, \frac{\lambda}{1-\frac{\lambda^2}{2}}\; \abs{\bar{\rho} - i\bar{\eta}} \, ,  
\qquad\qquad\qquad
\abs{\frac{V_{td}}{V_{ts}}}\, =\, \frac{\lambda}{1-\frac{\lambda^2}{2}}\; \abs{1- \frac{\lambda^2}{2} - \bar{\rho} -i\bar{\eta}} \, . 
\end{equation}
With that we find\ $\abs{V_{ts}^*V_{tb}}  = 0.0420 \pm 0.0011$. The rest of the inputs used in the analysis are given in Table~\ref{tab:table1}.

\begin{table}[t]
\centering
{\renewcommand{\arraystretch}{1.3}
\begin{tabular}{|c | c c|} 
 \hline
 Parameter & Value & Comment  \\ 
 \hline
$f_{B_d^0}$  & $(192.0 \pm 4.3)$ MeV & \cite{Aoki:2016frl} \\
$f_{B_s^0}$ & $(228.4 \pm 3.7)$ MeV & \cite{Aoki:2016frl}  \\
$f_{K} $ & $(155.6 \pm 0.4)$ MeV & \cite{Patrignani:2016xqp} \\ 
\hline
$\tau_{B_d^0}$ & $(1.520 \pm 0.004)$ ps & \cite{Amhis:2016xyh}\\
$\tau_{B_s}$& $(1.505 \pm 0.005)$ ps & \cite{Amhis:2016xyh}\\
$\frac{1}{\Gamma^s_H}$ & $(1.609 \pm 0.010)$ ps & \cite{Amhis:2016xyh}\\
$\frac{1}{\Gamma^s_L}$ & $(1.413 \pm 0.006)$ ps & \cite{Amhis:2016xyh}\\
$\Delta \Gamma_s $ & $(0.086\pm 0.006) \: \text{ps}^{-1}$ &\cite{Amhis:2016xyh}
\\
$\Delta m_{B_d^0}$ & $(0.5064\pm 0.0019)\:\text{ps}^{-1}$ &   \cite{Amhis:2016xyh}
\\
$\Delta m_{B_s^0}$ & $(17.757\pm 0.021)\:\text{ps}^{-1}$ &   \cite{Amhis:2016xyh}
\\
$m_t(m_t)$ & $(165.9\pm 2.1)$~GeV & \cite{Fuster:2017rev,Aad:2015waa} \\
\hline
$\abs{V_{ud}}$ &  $0.97417 \pm  0.00021$ & \cite{Hardy:2014qxa} \\
$\lambda$ & $0.2258 \pm 0.0009 $ & $(1-\abs{V_{ud}}^2)^{1/2}$\\
$\abs{V_{ub}}$ &  $(3.98 \pm 0.41) \cdot 10^{-3}$ &   \cite{Amhis:2016xyh} \\
$\abs{V_{cb}}$ & $(42.00 \pm 0.65) \cdot 10^{-3}$ & \cite{Gambino:2016jkc} 
\\
$A$ & $0.824 \pm 0.019$ & From $V_{cb}$ and $\lambda$ \\
$\bar{\rho}$ & $0.170 \pm 0.002$ & Our fit \\
$\bar{\eta}$ & $0.377 \pm 0.005$ & Our fit \\ 
\hline
$\mathrm{Br}(B_s^0 \rightarrow \mu^+ \mu^-)$ & $(3.0 \pm 0.6^{+0.3}_{-0.2}) \cdot 10^{-9} $ &  \cite{Aaij:2017vad} \\
$\mathrm{Br}(B_d^0 \rightarrow \mu^+ \mu^-)$ & $(1.5^{+1.2 \, +0.2}_{-1.0 \, -0.1})\cdot 10^{-10} $ &  \cite{Aaij:2017vad} \\
\hline
\end{tabular}
}
\caption{Inputs used in our analysis. Other masses and constants are taken from Ref.~\cite{Patrignani:2016xqp}.}
\label{tab:table1}
\end{table}

\section{$B_s^0 \rightarrow \mu^+ \mu^-$}
\label{sec:Bs}

A complete one-loop calculation of the $B_{d,s}^0 \rightarrow  \ell^+ \ell^-$ decay amplitudes within the A2HDM was performed in Ref.~\cite{Li:2014fea}\footnote{
The one-loop computation has been recently checked within (softly-broken) $\cZ_2$ models~\cite{Mescia}. The two calculations are in good agreement, except for a small difference in the $Z$-penguin contribution to $C_P$ which is numerically insignificant and originates in a different matching prescription.
},
including the effective one-loop FCNC local interaction of Eq.~\eqn{eq:FCNC2}, which is needed to properly reabsorb the UV divergences. The phenomenological study needs to be updated in view of the more precise LHCb measurement  \cite{Aaij:2017vad} of the time-integrated $B_s^0 \rightarrow  \mu^+ \mu^-$ branching ratio. Moreover, in Ref.~\cite{Li:2014fea} $\cC_d(\mu)$ was taken to be zero at $\mu = M_W$, in order to simplify the numerical analysis, while we are now interested in finding out how large this parameter could be. The decay $B_d^0 \rightarrow  \mu^+ \mu^-$ is also sensitive to the A2HDM contributions, but it leads to much weaker constraints at present, so we will concentrate in the $B_s^0$ decay mode. 

At the $B_q^0$ meson mass scale, the decay $B_q^0 \rightarrow  \ell^+ \ell^-$ can be described with the effective low-energy Hamiltonian
\beq
\cH_{\mathrm{eff}}\, =\, -\frac{G_F\alpha}{\sqrt{2}\pi\sin^2{\theta_W}}\;
V_{tb}^{\phantom{*}} V_{tq}^*\; \left\{ C_{10}\, O_{10} + C_S\, O_S + C_P\, O_P\right\}\, ,
\eeq
where 
\beq
O_{10}\, =\, (\bar q \gamma_\mu \cP_L b) (\bar\ell \gamma^\mu\gamma_5 \ell)\, ,
\qquad
O_S\, =\, \frac{m_b m_\ell}{M_W^2}\, (\bar q \cP_R b) (\bar\ell  \ell)\, ,
\qquad
O_P\, =\, \frac{m_b m_\ell}{M_W^2}\, (\bar q \cP_R b) (\bar\ell \gamma_5 \ell)\, ,
\eeq
with $m_b = m_b(\mu)$ the running $b$-quark mass and $\cP_{L/R} = (1\mp\gamma_5)/2$ the chirality projectors. Operators with the opposite quark chiralities are neglected because their contributions are very suppressed, being proportional to the light-quark mass $m_q$.

In the SM the scalar and pseudo-scalar Wilson coefficients are so tiny, that only the operator $O_{10}$ is numerically relevant. However $C_S$ and $C_P$ can be much more sizeable in models with extended scalar sectors. Neglecting any additional sources of CP violation beyond the CKM phase, the time-integrated branching ratio can be written as
\beq\label{eq:Br2m}
\overline{\mathcal{B}}(B_{q}^0 \rightarrow  \ell^+ \ell^-)\, =\,
\overline{\mathcal{B}}(B_{q}^0 \rightarrow  \ell^+ \ell^-)_{\mathrm{SM}}\;
\left\{ |P|^2 + \left( 1 -\frac{\Delta\Gamma_q}{\Gamma^q_L}\right)\, |S|^2
\right\}\, ,
\eeq
where
\begin{align}
 P &\equiv\, \frac{C_{10}}{C^{\rm SM}_{10}} + \frac{M^2_{B_q}}{2M^2_W} \left(\frac{m_b}{m_b+m_q}\right)\,\frac{C_P-C_P^{\mathrm{SM}}}{C^{\rm SM}_{10}} 
 \,,
 \label{eq:P}\\[0.2cm]
 S &\equiv\, \sqrt{1-\frac{4m^2_\ell}{M^2_{B_q}}}\; \frac{M^2_{B_q}}{2M^2_W} \left(\frac{m_b}{m_b+m_q}\right)\,\frac{C_S-C_S^{\mathrm{SM}}}{C^{\rm SM}_{10}} 
\,.
 \label{eq:S}
\end{align}
Complete analytical expressions for $C_{10}$, $C_P$ and $C_S$ are given in Ref.~\cite{Li:2014fea}. In the CP-conserving limit, they depend on ten A2HDM parameters: 3 Yukawa alignment factors ($\varsigma_u, \varsigma_d, \varsigma_\ell$), 3 scalar masses ($M_H, M_A, M_{H^{\pm}}$), 2 scalar potential couplings ($\lambda_3,\lambda_7$), the mixing angle $\tilde\alpha$ and the misalignment coefficient $\cC_d(M_W)$.

The only new-physics contribution to $C_{10}$ comes from $Z$-penguin diagrams
($Z$ exchange between the leptonic current and an effective $\bar q b Z$ vertex generated through one-loop diagrams with internal $H^\pm$ propagators):
\begin{equation} \label{eq:C10A2HDM}
\Delta C^{\rm A2HDM}_{10}\; =\; |\varsigma_u|^2\,\frac{x_t^2}{8}\, \left[\frac{1}{x_{H^+}-x_t} + \frac{x_{H^+}}{(x_{H^+}-x_t)^2}\,\left(\ln x_t - \ln x_{H^+}\right)\right]\,.
\end{equation}
It only depends on $|\varsigma_u|^2$ and the mass ratios $x_t\equiv m_t^2/M_W^2$ and $x_{H^+}\equiv M_{H^\pm}^2/M_W^2$.

The neutral scalar exchanges contribute to the scalar and pseudo-scalar Wilson coefficients. In the CP-conserving limit:
\beqn
\Delta C_S^{\varphi_i^0,\, \rm A2HDM} & = &
\frac{x_t}{2 x_h}\,\left( c_{\tilde{\alpha}} + s_{\tilde{\alpha}}\,\varsigma_\ell\right)\,
\Biggl\{s_{\tilde{\alpha}}\, (\varsigma_u-\varsigma_d)\, (1+\varsigma_u\,\varsigma_d)\, \cC_d(M_W)
\no\\ &&\hskip 3.0cm\mbox{}
+ \left( c_{\tilde{\alpha}}\, \lambda_3 + s_{\tilde{\alpha}}\, \lambda_7 \right)\,
\frac{2 v^2}{M_W^2}\; g_0^{\phantom{()}}+ c_{\tilde{\alpha}}\; g_1^{(a)} + s_{\tilde{\alpha}}\; g_2^{(a)}\Biggr\}
\no\\ &+&
\frac{x_t}{2 x_{H}}\,\left( c_{\tilde{\alpha}}\,\varsigma_\ell -s_{\tilde{\alpha}}\right)\,
\Biggl\{ c_{\tilde{\alpha}}\,  (\varsigma_u-\varsigma_d)\, (1+\varsigma_u\,\varsigma_d)\, \cC_d(M_W)
\label{eq:SexchCPC-S}\\ &&\hskip 3.0cm\mbox{}
- \left( s_{\tilde{\alpha}}\, \lambda_3 - c_{\tilde{\alpha}}\, \lambda_7 \right)\,
\frac{2 v^2}{M_W^2}\; g_0^{\phantom{()}}- s_{\tilde{\alpha}}\; g_1^{(a)} + c_{\tilde{\alpha}}\; g_2^{(a)}\Biggr\}\, ,\no
\\[0.2cm]
\Delta C_P^{\varphi_i^0,\, \rm A2HDM} & = &
-\varsigma_\ell\;\frac{x_t}{2 x_A}\;
\left[ (\varsigma_u-\varsigma_d)\,(1+\varsigma_u\,\varsigma_d)\, \cC_d(M_W) + g_3^{(a)}\right]\, ,
\label{eq:SexchCPC-P}
\eeqn
where $c_{\tilde{\alpha}}=\cos\tilde{\alpha}$ and $s_{\tilde{\alpha}}=\sin\tilde{\alpha}$ are the scalar mixing factors, and $x_{\varphi_i^0}\equiv M_{\varphi_i^0}^2/M_W^2$  with $\varphi_i^0 = h, H, A$. 
The functions $g_0(x_t,x_{H^+},\varsigma_u,\varsigma_d)$ and $g_i^{(a)}(x_t,x_{H^+},\varsigma_u,\varsigma_d)$ ($i=1,2,3$) can be found in the appendix of 
Ref.~\cite{Li:2014fea}. We do not reproduce them here to avoid reiterating lengthy formulae. There are, in addition, box-diagram contributions to $C_{S,P}$
and $Z$-penguin contributions to $C_P$, which only depend on the three alignment parameters $\varsigma_f$ and the mass ratios $x_t$ and $x_{H^+}$; their explicit expressions are also given in Ref.~\cite{Li:2014fea}.\footnote{
All gauge-dependent terms have been removed from \eqn{eq:SexchCPC-S} and \eqn{eq:SexchCPC-P} since they must be combined with boxes and $Z$-penguin diagrams to get gauge-independent results. See Ref.~\cite{Li:2014fea} for details.}
The SM Higgs-exchange contribution can be easily recovered from Eq.~\eqn{eq:SexchCPC-S} by taking the appropriate limit: $\varsigma_f, s_{\tilde\alpha},\lambda_{3,7}\to 0$, $x_{H,H^+}\to\infty$.

Once constrained in the range $\cos{\tilde\alpha}\in [0.9, 1]$, the mixing angle has a very marginal impact on the predictions. Therefore, we will choose   
$\cos{\tilde\alpha} = 0.95 $ to simplify the numerical analysis.
Since the results are not very sensitive either to the scalar potential parameters, we will also set $\lambda_3 = \lambda_7 = 1$.\footnote{By varying $\lambda_{3,7}$ in the perturbative allowed region the ratio $\frac{\mathrm{Br}(B_s^0 \rightarrow \mu^+ \mu^-)}{\mathrm{Br}(B_s^0 \rightarrow \mu^+ \mu^-)_{\mathrm{SM}}}$ varies in less than a 1\%.} 
The current (95\% CL) experimental constraints on $\cC_d(M_W)$ are displayed in Figs.~\ref{fig:Bs1}, \ref{fig:Bs2} and \ref{fig:Bs3}, for different choices of the remaining free parameters. The left and right panels on these three figures correspond to $\varsigma_\ell =0$ and $\varsigma_\ell =30$, respectively.
Fig.~\ref{fig:Bs1} exhibits the correlated constraints on the plane $\cC_d(M_W)$, $\varsigma_d$, taking $\varsigma_u = 0$. Fig.~\ref{fig:Bs2} shows the constraints on $\cC_d(M_W)$ and $\varsigma_u$, taking $\varsigma_d = 0$, while a large value 
$\varsigma_d = 50$ is adopted in Fig.~\ref{fig:Bs3}. Different assumptions on the scalar mass spectrum are analysed in all these figures. 

The plots take also into account the constraints enforced by the weak radiative decay $\bar{B} \rightarrow X_s \gamma$ \cite{Jung:2010ik,Jung:2010ab,Jung:2012vu,Misiak:2006ab, Hermann:2012fc, Bobeth:1999ww, Misiak:2006zs}, which drastically reduce the allowed parameter space, specially for large values of $\varsigma_u^*\varsigma_d$. The Wilson coefficients that are relevant for this process take the form
 $C_i^{\mathrm{eff}} = C_{i,\mathrm{SM}} + |\varsigma_u|^2 C_{i,uu} -(\varsigma_u^*\varsigma_d) C_{i,ud}$, where $C_{i,uu}$ and $C_{i,ud}$ contain the dominant A2HDM contributions from virtual top and $H^\pm$ propagators \cite{Jung:2010ik}. The combined result is very sensitive to the ratio $\varsigma_d/\varsigma_u$, implying a correlated constraint on $\varsigma_d$,  $\varsigma_u$ and $M_{H^\pm}$ that becomes very strong for real values of the alignment parameters. This constraint may be relaxed by including a (CP-violating) relative phase between $\varsigma_d$ and $\varsigma_u$ \cite{Jung:2010ik,Jung:2010ab,Jung:2012vu}.

\begin{figure}[t] 
\centering
\subfloat{
\includegraphics[scale=0.5]{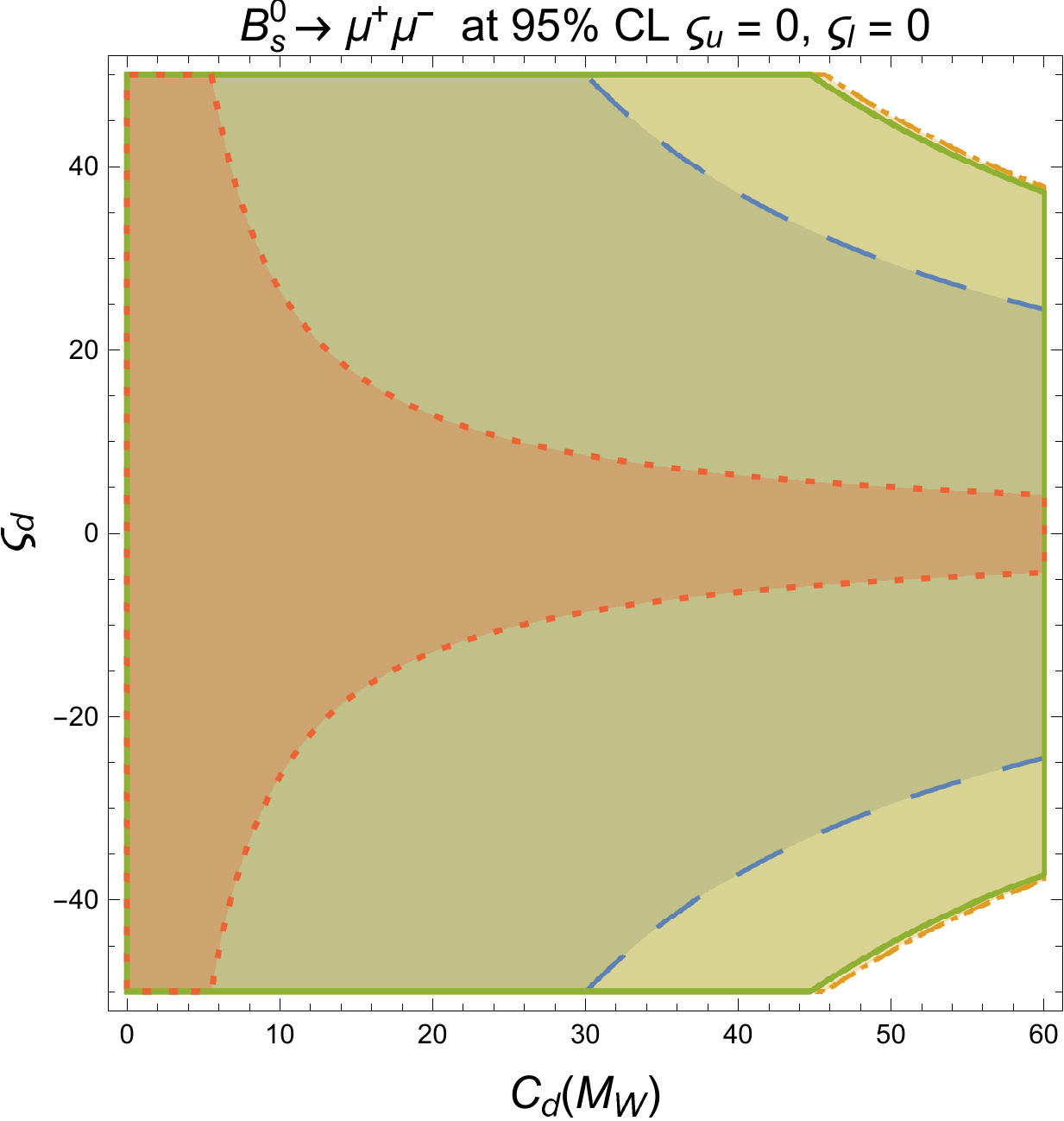}
}
\hskip 1.75cm
\subfloat{
\includegraphics[scale=0.5]{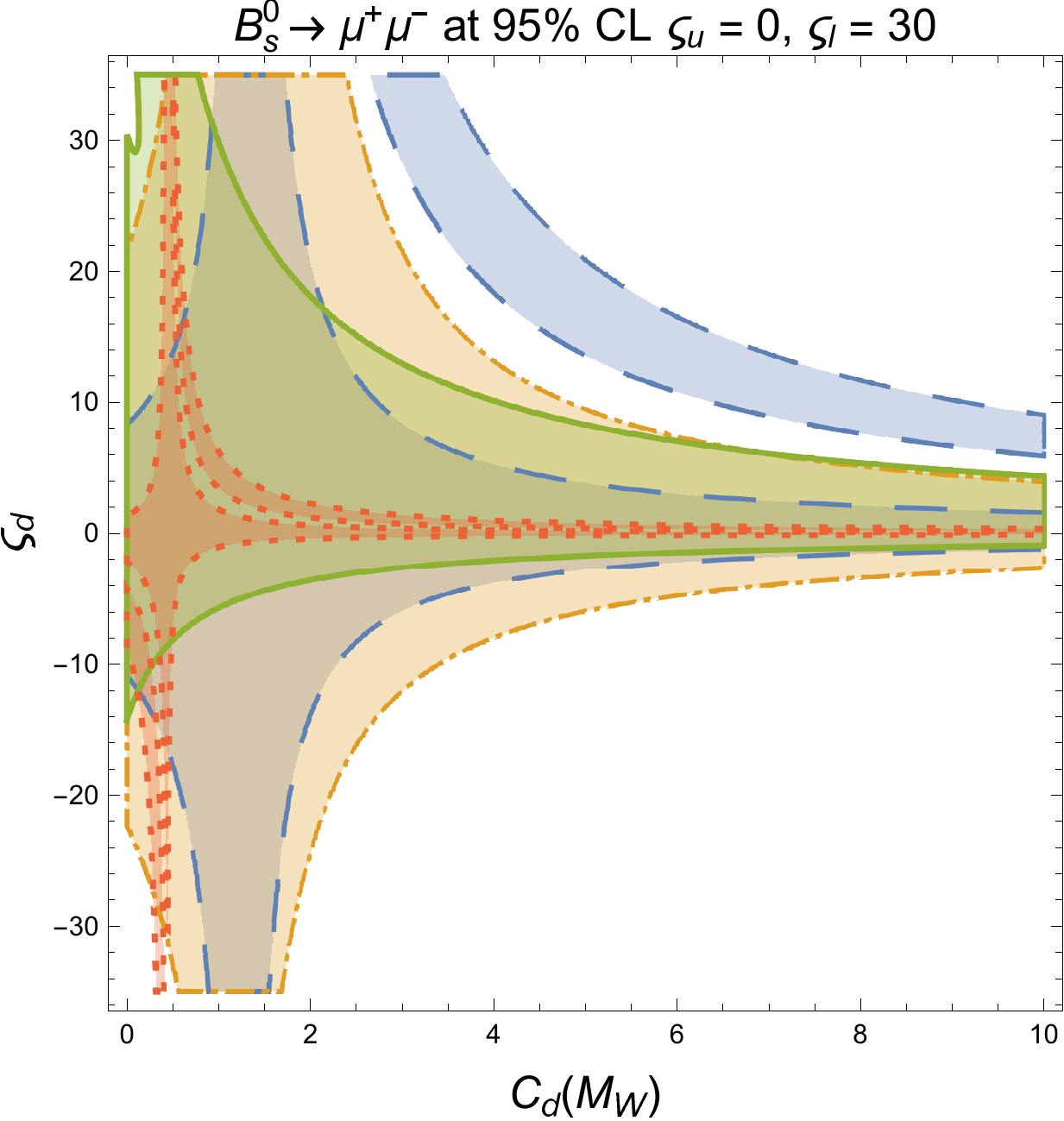}
}
\caption{$B_s^0 \rightarrow  \mu^+ \mu^-$ constraints on $\cC_d(M_W)$ and $\varsigma_d$, in the CP-conserving limit, for $\lambda_3 = \lambda_7 =  1$, $c_{\tilde\alpha} = 0.95$ and $\varsigma_u=0$, with $\varsigma_l=0$ (left) and $\varsigma_l=30$ (right). The coloured areas show the allowed regions (95\% CL) for different mass configurations defined in Eq.~\eqref{eq:massconf}: A (red, dotted), B (green, solid line), C (blue, dashed) and D (orange, dot-dashed).}
\label{fig:Bs1}
\end{figure}

\begin{figure}[th]
\centering
\subfloat{
\includegraphics[scale=0.5]{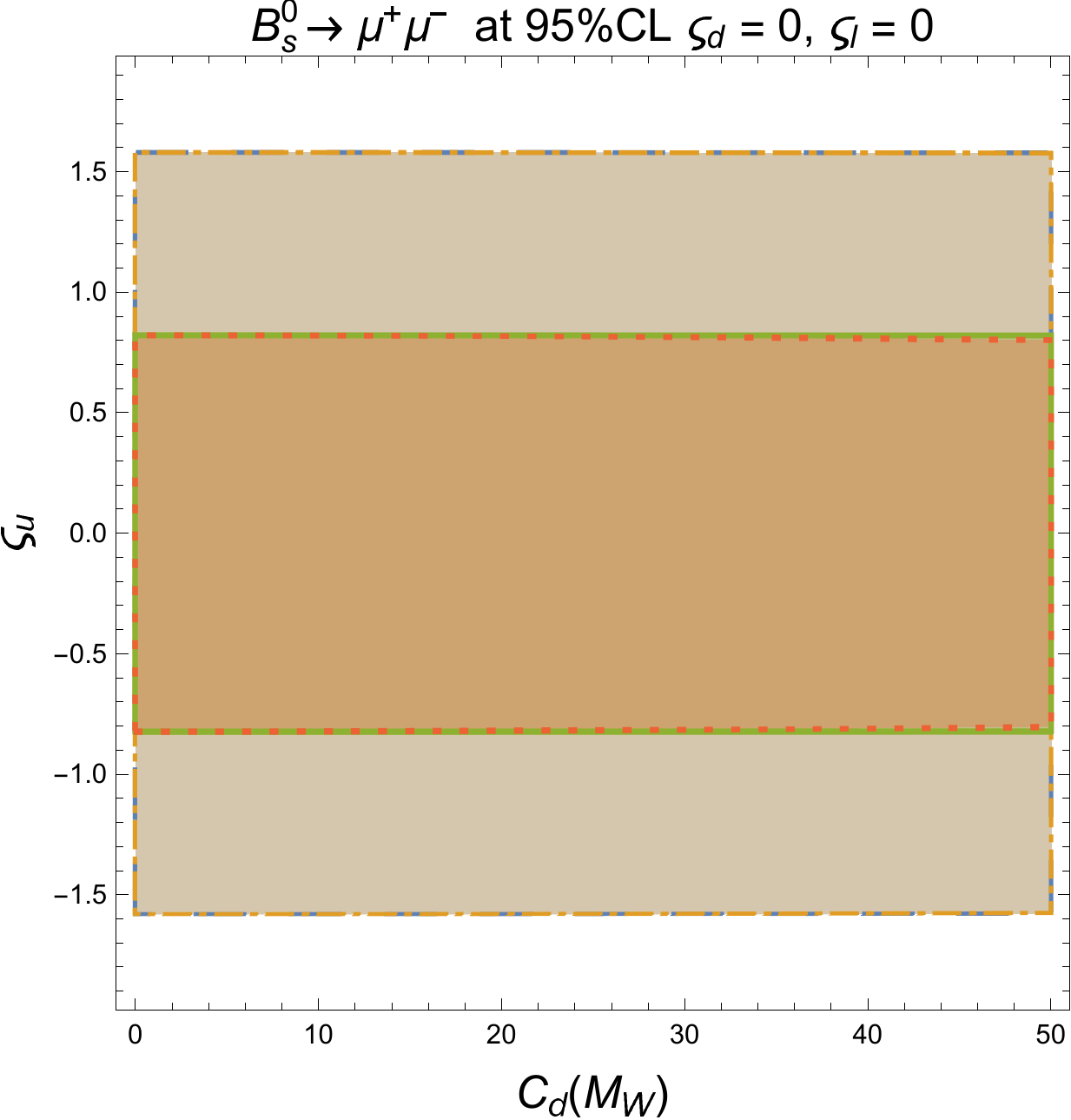}
}
\hskip 1.75cm
\subfloat{
\includegraphics[scale=0.5]{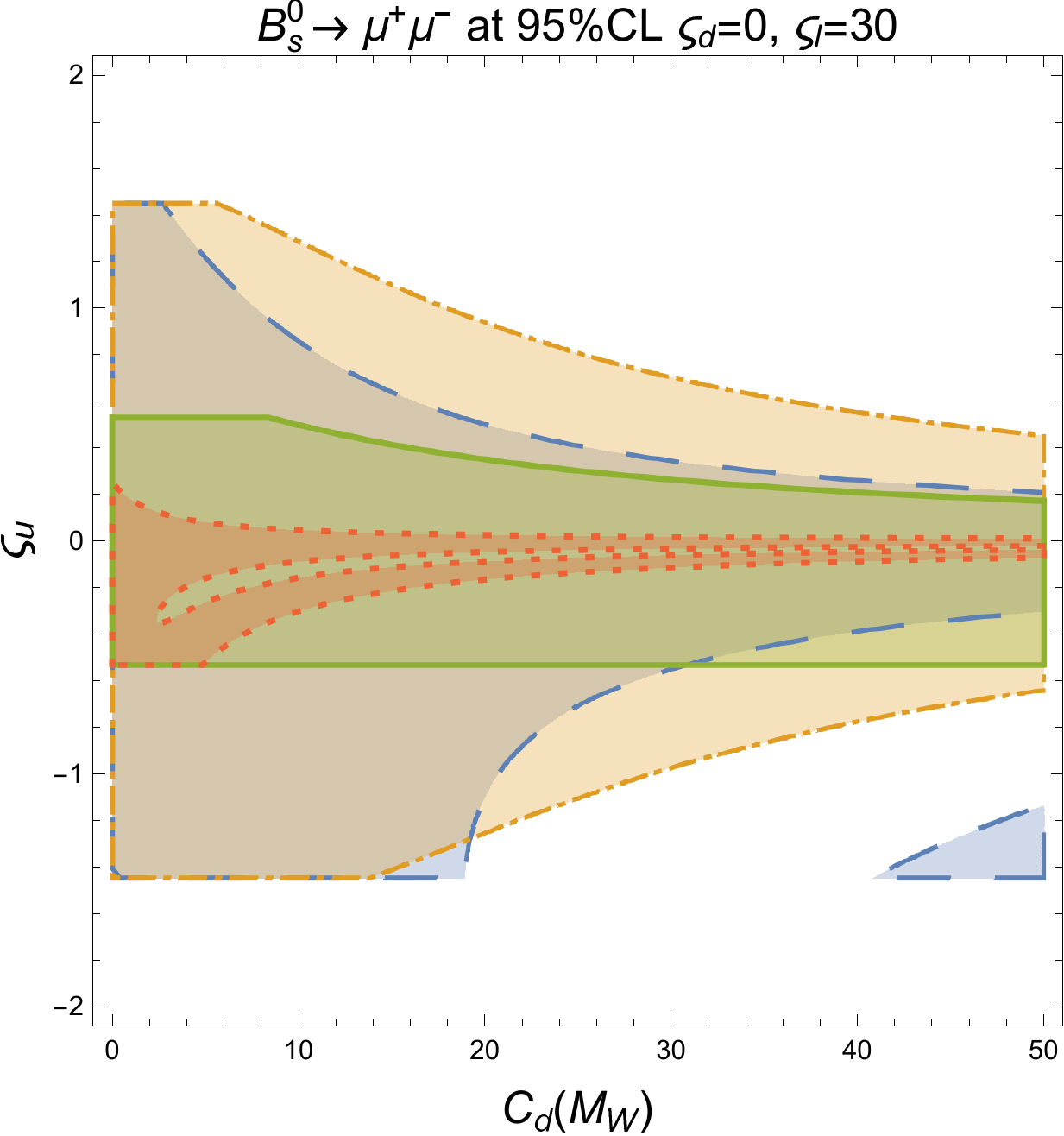}
}
\caption{$B_s^0 \rightarrow  \mu^+ \mu^-$ constraints (95\% CL) on $\cC_d(M_W)$ and $\varsigma_u$, in the CP-conserving limit, for $\lambda_3 = \lambda_7 =  1$, $c_{\tilde\alpha} = 0.95$ and $\varsigma_d=0$, with $\varsigma_l=0$ (left) and $\varsigma_l=30$ (right). Same colour coding than Fig.~\ref{fig:Bs1}.}
\label{fig:Bs2}
\end{figure}

\begin{figure} 
\centering
\subfloat{\noindent
\includegraphics[scale=0.42]{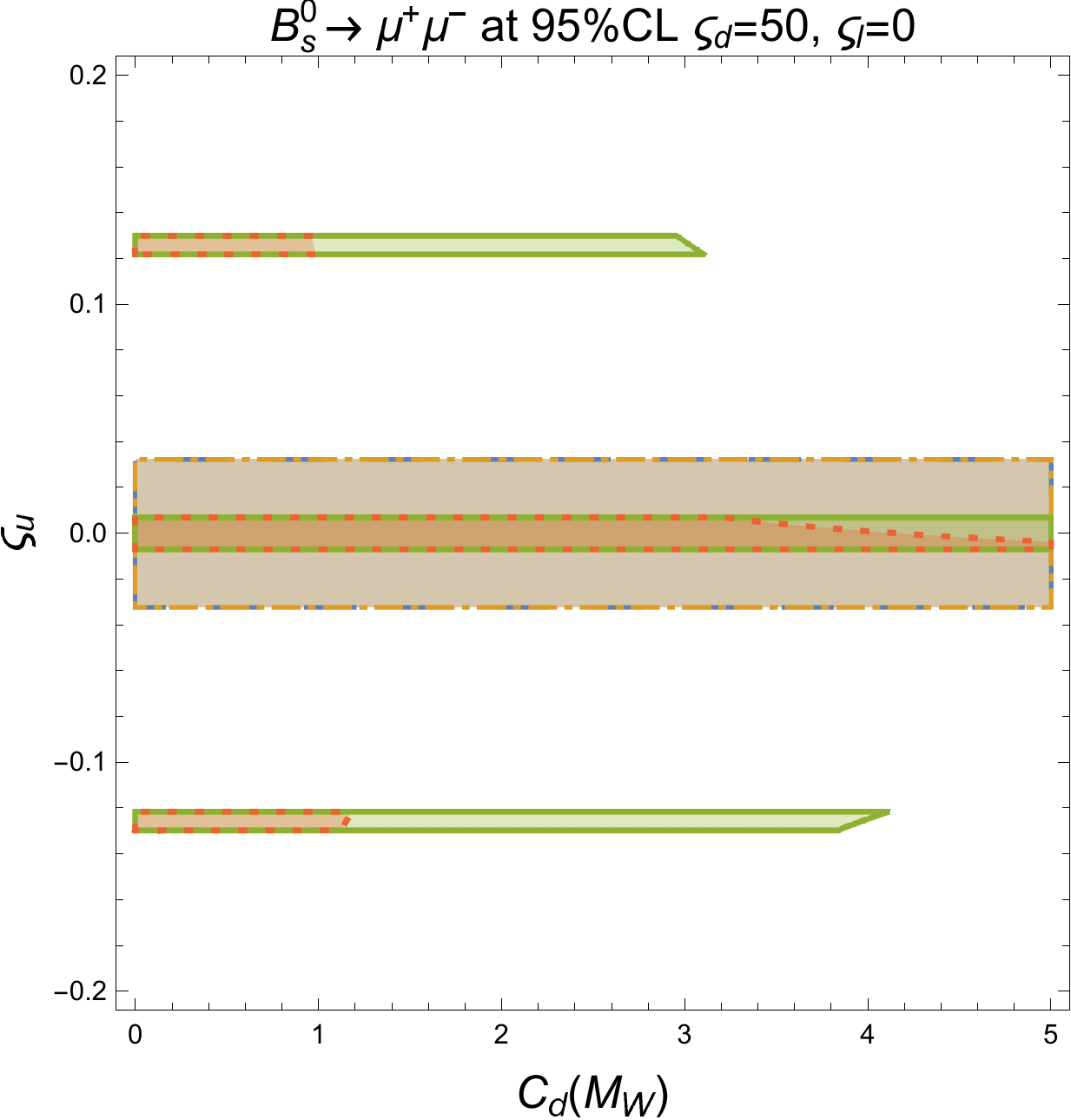}
}
\subfloat{
\includegraphics[scale=0.42]{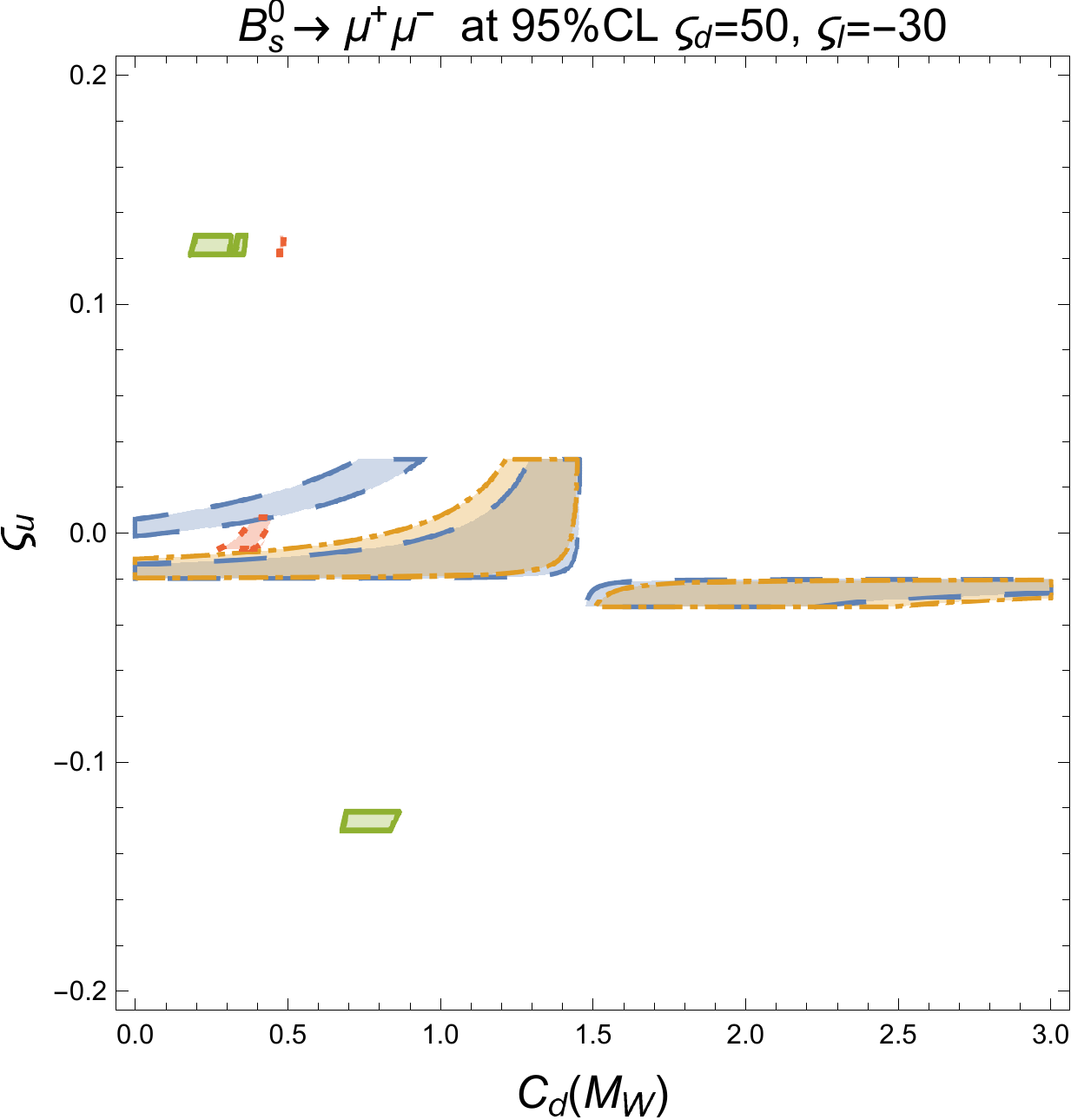}
}
\subfloat{
\includegraphics[scale=0.42]{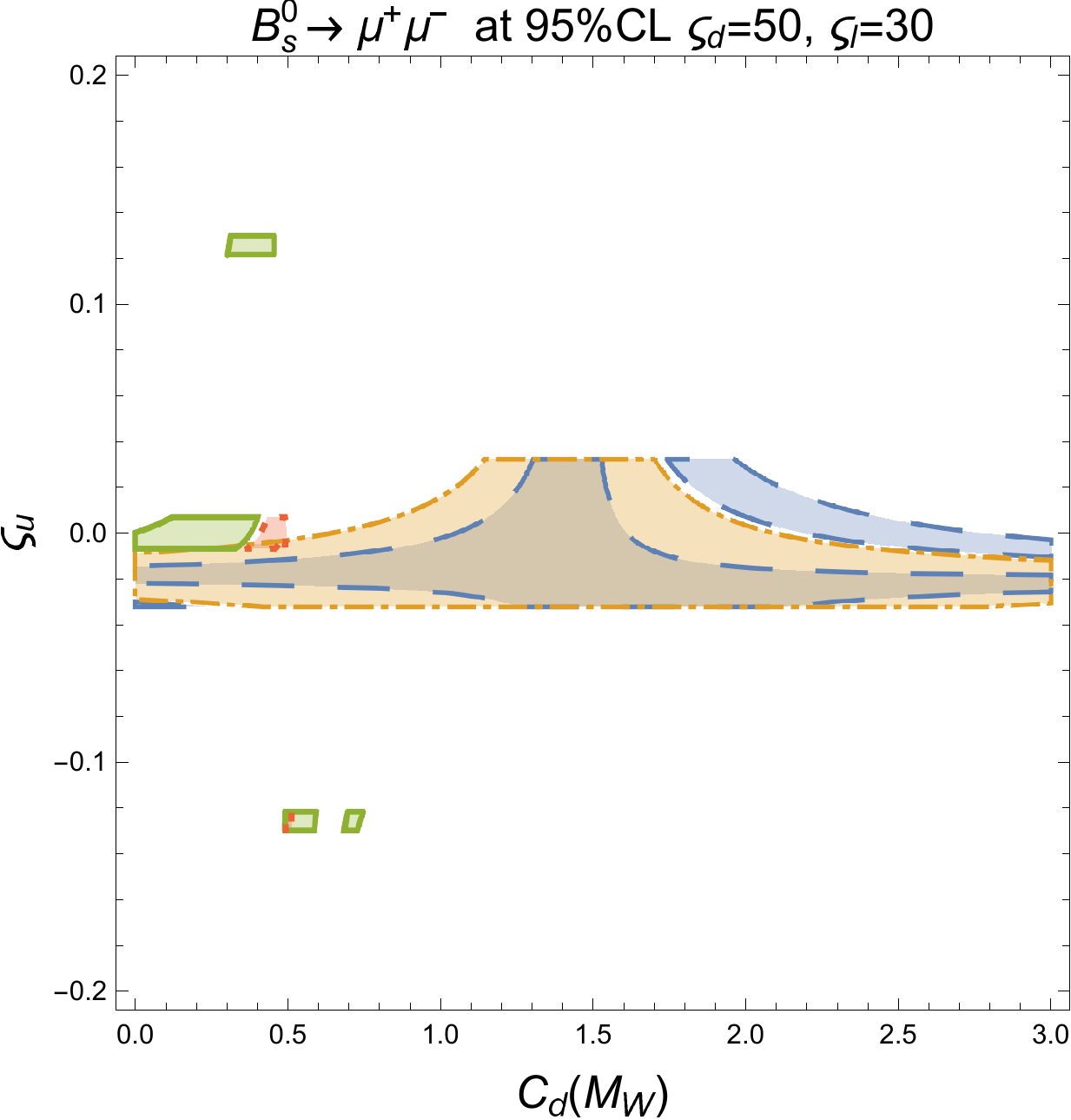}
}
\caption{$B_s^0 \rightarrow  \mu^+ \mu^-$ constraints (95\% CL) on $\cC_d(M_W)$ and $\varsigma_u$, in the CP-conserving limit, for $\lambda_3 = \lambda_7 =  1$, $c_{\tilde\alpha} = 0.95$ and $\varsigma_d=50$, with $\varsigma_l=0$ (left), $\varsigma_l=-30$ (middle) and $\varsigma_l=+30$ (right) . Same colour coding than Fig.~\ref{fig:Bs1}.}
\label{fig:Bs3}
\end{figure}

\begin{figure}[t]
\centering
\subfloat{
\includegraphics[scale=0.5]{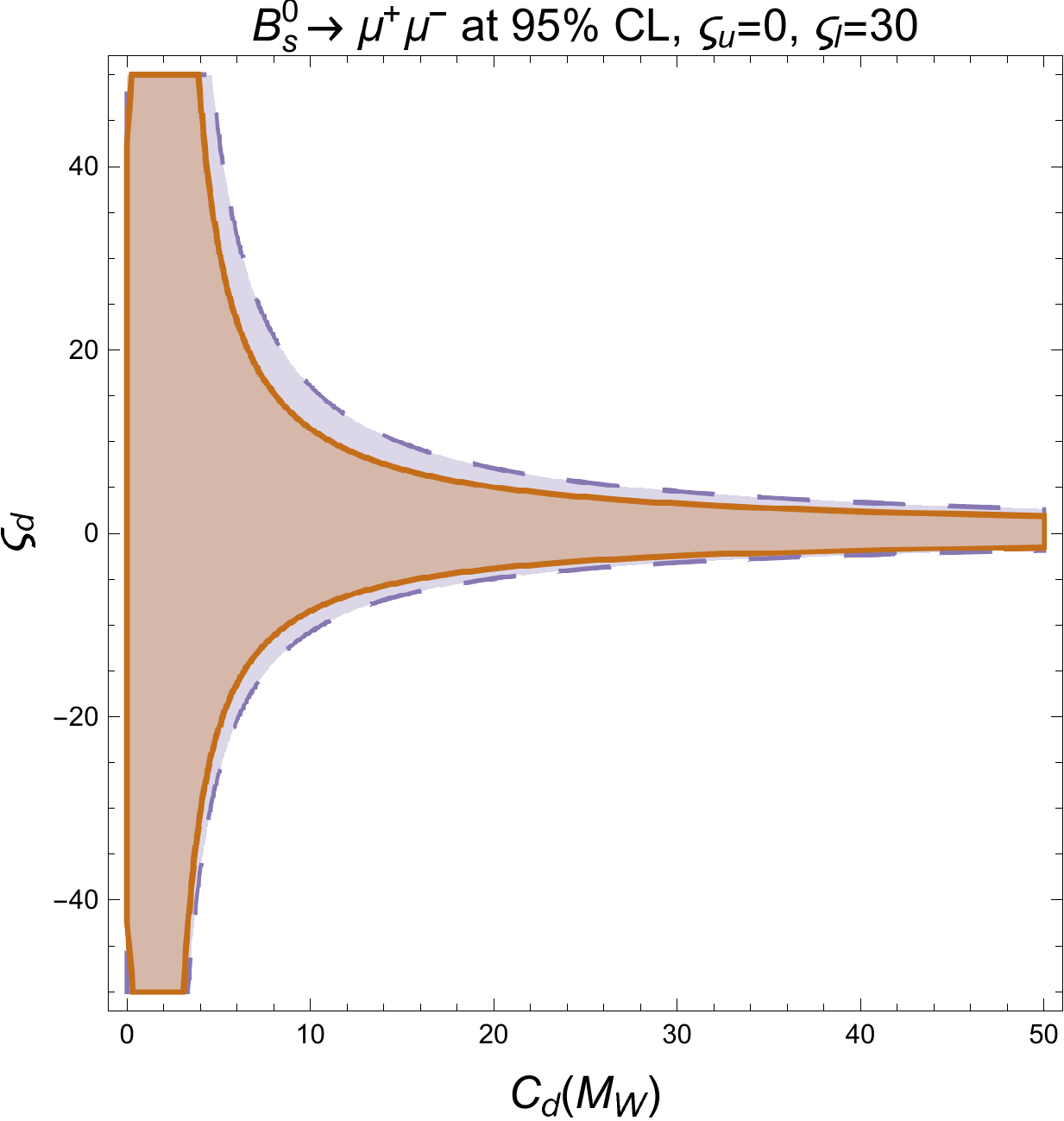}
}
\hskip 1.75cm
\subfloat{
\includegraphics[scale=0.5]{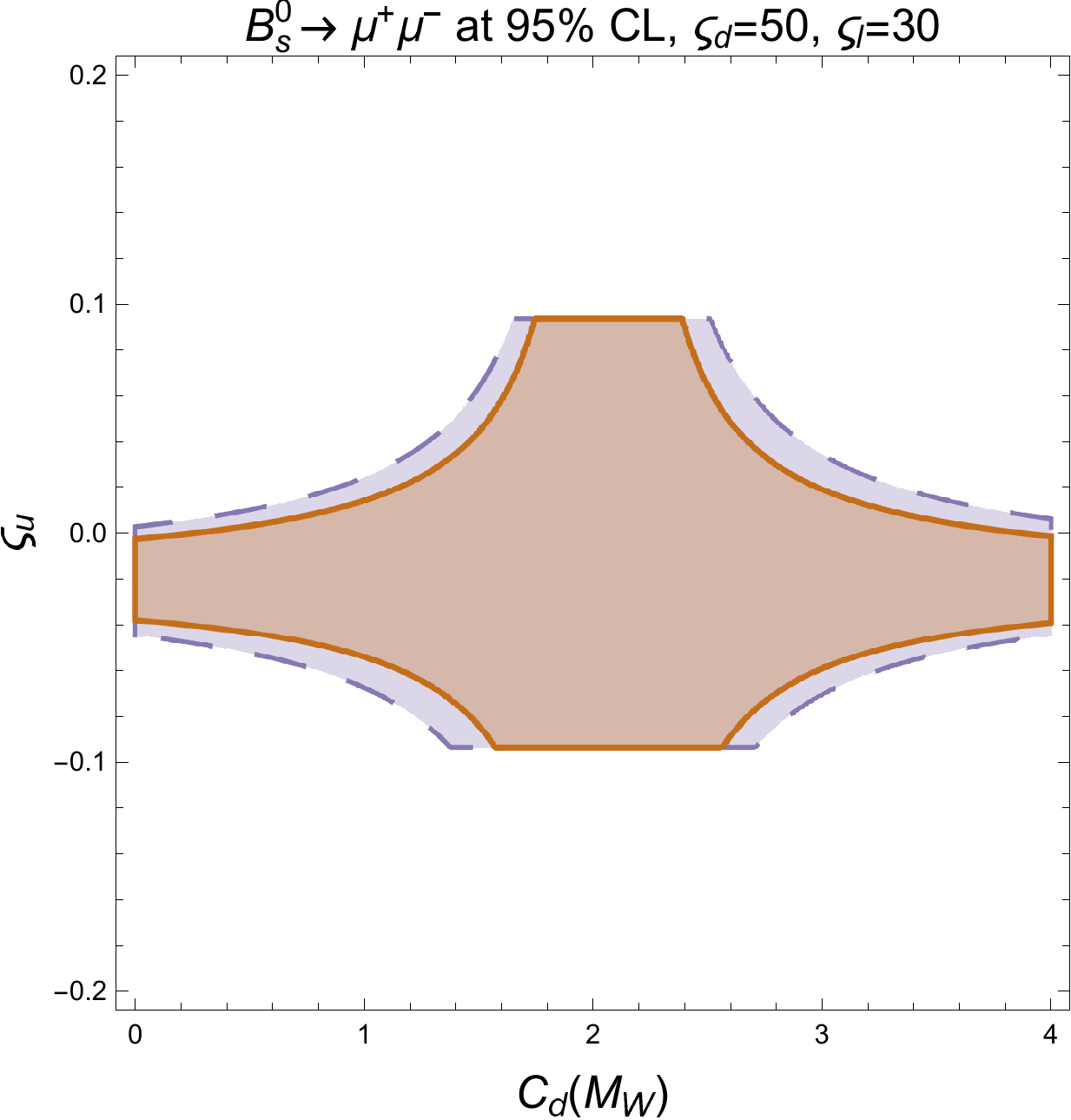}
}
\caption{The left (right) panel shows the
$B_s^0 \rightarrow  \mu^+ \mu^-$ constraints (95\% CL) on $\cC_d(M_W)$ and $\varsigma_d$ ($\varsigma_u$), in the CP-conserving limit, for $\lambda_3 = \lambda_7 =  1$, $c_{\tilde\alpha} = 0.95$ and $\varsigma_l=30$, with $\varsigma_u=0$ ($\varsigma_d = 50$) and the two heavy-mass configurations in Eq.~\eqref{eq:massconf}: E (orange, solid line) and F (violet, dashed).}
\label{fig:extra}
\end{figure}


The following generic conclusions can be extracted:

\begin{itemize}
\item Since the misalignment contribution is proportional to $(\varsigma_u-\varsigma_d) (1+\varsigma_u\varsigma_d)$, there are no constraints on $C_d(M_W)$ at $\varsigma_u=\varsigma_d$ or $\varsigma_u=-1/\varsigma_d$. These specific values of the alignment parameters correspond to models with natural flavour conservation, where $\cL_{\mathrm{FCNC}}=0$.

\item The comparison of the left and right panels shows the importance of the terms proportional to $\varsigma_\ell$. At $\varsigma_\ell=0$ many A2HDM contributions are eliminated: all box corrections with $H^\pm$ exchanges vanish in this limit
and all diagrams mediated through non-SM scalars are removed, 
up to small mixing effects proportional to $s_{\tilde\alpha}$; only the $Z$-penguin and the SM Higgs-exchange diagrams survive. 
$\Delta C_P^{\varphi_i^0,\, \rm A2HDM}$ vanishes identically at $\varsigma_\ell=0$, while the misalignment contribution to $C_S$ is proportional to
$c_{\tilde\alpha} s_{\tilde\alpha} (x_H-x_h)$, disappearing when the mixing angle or the neutral mass splitting approach zero. Therefore, if $\varsigma_\ell=0$, no constraints on $\cC_d(M_W)$ can be set at $c_{\tilde\alpha}=1$ or when $M_H=M_h$.

\item When $\varsigma_u=0$, there are no charged-scalar contributions to $\bar B\to X_s\gamma$. Therefore the constraints displayed in Fig.~\ref{fig:Bs1} and the left panel of Fig.~\ref{fig:extra} fully originate from the decay $B_s^0 \rightarrow  \mu^+ \mu^-$. Moreover, $\Delta C_{10}^{\mathrm{A2HDM}}\propto |\varsigma_u|^2 =0$, and the $Z$-penguin A2HDM correction to $C_P$ is also zero. The misalignment contributions to $C_{S,P}$ are proportional in this case to $\varsigma_d\, C_d(M_W)$, which explains the $C_d(M_W)\lsim 1/\varsigma_d$ scaling exhibited in Figs.~\ref{fig:Bs1} and \ref{fig:extra} (left). If additionally $\varsigma_\ell=\varsigma_u=0$, the only non-zero scalar contributions are $\Delta C_S^{h,\mathrm{A2HDM}}$ and $\Delta C_S^{H,\mathrm{A2HDM}}$, which are obviously independent of $M_A$ and generate the strong dependence on $M_H$, roughly scaling as $1/M_H^2$, displayed on Fig.~\ref{fig:Bs1} (left).
The right panel in Fig.~\ref{fig:Bs1} shows that much stronger constraints are obtained with $\varsigma_\ell\not=0$. The allowed regions obviously expand with increasing scalar masses. Notice, however, how the configurations A (red) and C (blue), with $M_A < M_{H^\pm}$, generate additional allowed bands, not present for B (green) and D (orange), which originate in
the interference of $\Delta C_P^{A,\mathrm{A2HDM}}$ with box-diagram contributions to $C_P$ proportional to the product $\varsigma_\ell\varsigma_d$.

\item For small values of $|\varsigma_{d,\ell}|\le |\varsigma_u|$, the one-loop contributions to $C_{S,P}$ are negligible compared to $\Delta C_{10}^{\mathrm{A2HDM}} \propto |\varsigma_u|^2$. The measured rate $\overline{\mathcal{B}}(B_{q}^0 \rightarrow  \mu^+ \mu^-)$ provides then an upper bound on $|\varsigma_u|$ that is stronger than the one extracted from $Z\to b\bar b$ and only depends on $M_{H^\pm}$ \cite{Li:2014fea}. As shown in the left panel of Fig.~\ref{fig:Bs2}, this limit (identical for configurations A and B, and also for C and D) is independent on $\cC_d(M_W)$. For very large values of $\cC_d(M_W)$, such that the misalignment contribution $\sim \varsigma_u\, \cC_d(M_W)$ could be sizeable, the upper bound on $|\varsigma_u|$ would obviously become stronger.

\item At large values of $\varsigma_\ell$, the misalignment contribution to $C_{S,P}$ increases proportionally to $\varsigma_\ell$. This needs to be compensated with smaller values of both $\varsigma_u$ and $\varsigma_d$, in order to satisfy the  
$\overline{\mathcal{B}}(B_{q}^0 \rightarrow  \mu^+ \mu^-)$ constraint. Thus, sizeable values of $\cC_d(M_W)$ imply very small quark alignment parameters. The figures show, however, that this can be avoided at very specific values of $\cC_d(M_W)$ where the misalignment and loop contributions cancel.

\item The restrictions imposed by $\bar{B} \rightarrow X_s \gamma$ can completely dominate over constraints coming from $B_{s}^0 \rightarrow \mu^+ \mu^-$ at large values of $\varsigma_d$. This is reflected in the horizontal bands in the left panel of Fig.~\ref{fig:Bs3}. The $B_{s}^0 \rightarrow \mu^+ \mu^-$ data puts nevertheless a limit on $|C_d(M_W)|$ for non-zero values of $\varsigma_u$. Allowing also for large values of $|\varsigma_\ell|$, the combined constraints from $\bar{B} \rightarrow X_s \gamma$ and 
$B_{s}^0 \rightarrow \mu^+ \mu^-$ become very stringent, as shown in the middle and right panels of Fig.~\ref{fig:Bs3}, which also illustrate the impact of the $\varsigma_d\varsigma_\ell$ sign.

\item  When the scalar masses are increased, the new-physics contributions gradually decouple and the allowed regions become larger. This is shown in Fig.~\ref{fig:extra}, taking $\varsigma_\ell = 30$ and two different mass configurations: E ($M_{H^{\pm}}= M_{A} = 10^3$ GeV, $M_{H} =500$ GeV; orange) and  F ($M_{H^{\pm}} = M_H = M_A = 10^3$ GeV; violet). Taking $M_{H^{\pm}}= M_{H} = 10^3$ GeV and $M_{A} =500$ GeV gives results similar to the E configuration. The left (right) panel shows the constraints on $C_d(M_W)$ and $\varsigma_d$ ($\varsigma_u$), for $\varsigma_u=0$ ($\varsigma_d=50$). They should be compared with the analogous plots for lighter mass configurations in the right panels of Figs.~\ref{fig:Bs1} and \ref{fig:Bs3}.

\end{itemize}

\section{Meson mixing}
\label{sec:mesonmixing}

As already commented before, two insertions of $\cL_{\mathrm{FCNC}}$ are needed in order to generate a misalignment contribution to meson-antimeson mixing. This is a two-loop correction and, therefore, it is expected to be quite small. Nevertheless, previous tree-level analyses of $\cL_{\mathrm{FCNC}}$ have focused on the $\Delta B=2$ transition, owing to the high sensitivity of $B^0_q$--$\bar B^0_q$ mixing to new-physics effects, 

The one-loop scalar contribution to the neutral meson mixing has been analysed, within the A2HDM, in Refs.~\cite{Jung:2010ik,Chang:2015rva,Cho:2017jym}. It proceeds through box diagrams with internal $H^\pm$ propagators and provides stringent constraints on $|\varsigma_u|$, which depend on $M_{H^\pm}$. Actually both the $B_s^0$--$\bar B_s^0$ mass difference and the CP-violating $\varepsilon_K$ parameter provide bounds on $|\varsigma_u|$ which are quite similar to the ones extracted from $Z\to b\bar b$ \cite{Jung:2010ik}. So far, we did not use this information because we would like to get constraints on $\cC_d$, which was not taken into account in those one-loop analyses.

While being a second-order effect, the neutral scalar exchange between two $\cL_{\mathrm{FCNC}}$ vertices could be of a similar size, or even larger, than the one-loop charged scalar contribution, due to a large $\cC_d$ coupling or a very light neutral scalar. However, the fact that the analyses of $\Delta M_{B^0_q}$ and $\varepsilon_K$, without any misalignment contribution, give similar constraints than $Z\to b\bar b$ does not seem to favour this possibility. This is also confirmed by our previous study of $B_s^0\to\mu^+\mu^-$, although the constraints on $\cC_d$ obtained there could be avoided for some specific choices of A2HDM parameters (for instance, $\varsigma_\ell = s_{\tilde\alpha} = 0$). 

The (one-loop) charged-current and (tree-level) misalignment contributions to $B_q^0$--$\bar B_q^0$ mixing are roughly proportional to the factors
\beq
\omega_{\mathrm{CC}}\, =\, \frac{1}{16\pi^2}\,\frac{m_t^4}{M_{H^\pm}^2 v^4}\; \left(V_{tq}^* V_{tb}^{\phantom{*}}\right)^2\, ,
\qquad\qquad\qquad
\omega_{\mathrm{NC}}\, =\, \frac{|\cC_d(\mu)|^2}{16\pi^4}\,\frac{m_b^2 m_t^4}{M_{\varphi^0_k}^2 v^6}\; \left(V_{tq}^* V_{tb}^{\phantom{*}}\right)^2\, .
\eeq
Their relative size scales approximately as $\omega_{\mathrm{NC}}/\omega_{\mathrm{CC}} = |\cC_d(\mu)|^2 m_b^2 M_{H^\pm}^2/(M_{\varphi^0_k}^2 v^2\pi^2)$.
In order to have a ratio $\omega_{\mathrm{NC}}/\omega_{\mathrm{CC}}\sim \cO(1)$,
one needs $|\cC_d(\mu)| M_{H^\pm}/M_{\varphi^0_k}\sim \cO(10^2)$. 
A proper calculation of the misalignment effects would require in any case the inclusion of two-loop diagrams in order to cancel the renormalization-scale dependence of  $\cC_d(\mu)$.\footnote{ In the absence of a complete two-loop computation, one could extract effective $\mu$-independent $\varphi^0_k\bar q b$ vertices from the $B_q^0\to\ell^+\ell^-$ computation presented in Ref.~\cite{Li:2014fea}. However, they would still contain small gauge dependences.}

To estimate the possible size of the misalignment correction, we will  consider the tree-level scalar exchange in Fig.~\ref{fig:diagrams} (right),  taking $\mu=M_W$ to normalize the coupling $\cC_d$. It contributes to the effective low-energy Hamiltonian,
\begin{equation} \label{eq:Heff}
\mathcal{H}_{\mathrm{eff}}\, \supset \, \sum_{i,j=d,s,b} \left\{ C_{1,ij}^{SRR}\, \mathcal{O}_{1,ij}^{SRR} + C_{1,ij}^{SLL}\, \mathcal{O}_{1,ij}^{SLL} + C_{2,ij}^{LR}\, \mathcal{O}_{2,ij}^{LR} \right\}\, ,
\end{equation}
generating $\Delta S = 2$ and $\Delta B = 2$ transitions through the four-quark operators
\beqn\label{eq:DF2op}
\mathcal{O}_{1,ij}^{SRR} \, =\, (\bar{d}_{i L} d_{j R}) (\bar{d}_{i L} d_{j R})\, , \quad\;
\mathcal{O}_{1,ij}^{SLL} \, =\, (\bar{d}_{i R} d_{j L}) (\bar{d}_{i R} d_{j L})\, ,  
\quad\;\mathcal{O}_{2,ij}^{LR} \, =\, (\bar{d}_{i R} d_{j L}) (\bar{d}_{i L} d_{j R}) \, , 
\eeqn
with
\beqn\label{eq:WilsonC}
C_{1,ij}^{SRR} \, =\, \frac{g_{ij}^2}{16 \pi^4 v^6}\,  \sum_{k=1}^3 E_k^2\, , 
\qquad
C_{1,ij}^{SLL} \, =\, \frac{g_{ji}^{*2}}{16 \pi^4 v^6}\,  \sum_{k=1}^3 E_k^{*2}\,  , 
\qquad
C_{2,ij}^{LR} \, =\, \frac{g_{ij} g_{ji}^{*}}{8 \pi^4 v^6} \,  \sum_{k=1}^3 |E_k|^2 \, .
\eeqn
To simplify the numerical analysis, we have split the Wilson coefficients into a
global constant that reabsorbs all A2HDM parameters,
\beqn
E_k \,\equiv\, \cC_d(M_W) (\varsigma_d- \varsigma_u) (1 + \varsigma_d \varsigma_u^*)\;\frac{1}{M_{\varphi_k^0}}\,
\left( \mathcal{R}_{k2} + i\, \mathcal{R}_{k3}\right)\, ,
\eeqn
and a flavour structure which is fully determined by the quark masses and mixings,
\beqn
g_{ij} \,\equiv\, \left( V_{\text{CKM}}^{\dagger} M_u M_u^{\dagger} V_{\text{CKM}}^{\phantom{\dagger}} M_d \right)_{ij} \, .
\eeqn
Neglecting any additional source of CP violation beyond the CKM phase, $E_1$ and $E_2$ are real, while $E_3$ is imaginary; this implies different relative signs for the CP-even and CP-odd scalar contributions to $C_{1,ij}^{SRR}$ and $C_{1,ij}^{SLL}$, while they enter with the same sign in $C_{2,ij}^{LR}$.

In our phenomenological analysis we have also included the full one-loop charged-current contribution \cite{Jung:2010ik,Chang:2015rva,Cho:2017jym}, which is obviously $\mu$-independent. The hadronic matrix elements of the $\Delta F = 2$ four-quark operators \eqn{eq:DF2op} are detailed in appendix~\ref{app:Bpar}. The most restrictive limits are obtained from $B_s^0$--$\bar B_s^0$ mixing (slightly weaker bounds result from $B_d^0$--$\bar B_d^0$ mixing and $\varepsilon_K$), taking always into account the correlated restrictions from $\bar B\to X_s\gamma$. The measured mass difference in the $B_s^0$--$\bar B_s^0$ system imposes stringent constraints on $\varsigma_u$, $\varsigma_d$ and $M_{H^\pm}$, originating in the one-loop contributions, but the sensitivity to the misalignment parameter is quite small, except at very large values of $|\varsigma_{d}|$. This is illustrated in Fig.~\ref{fig:ConstraintsMeson} which shows two different parametric configurations, $\varsigma_d=50$ (left) and $\varsigma_u=0.5$ (right). In both cases one observes horizontal lines, exhibiting the low sensitivity to $C_d(M_W)$. Nevertheless, a bound on $C_d(M_W)$ finally emerges when $\varsigma_d\, C_d(M_W)$ is large enough to generate a sizeable misalignment effect. The panels display the same mass configurations analysed in the previous section (C and D give here equivalent results). Obviously, the sensitivity to $C_d(M_W)$ is larger for low scalar masses (configurations A and B).

\begin{figure}
\centering
\subfloat{
\includegraphics[scale=0.5]{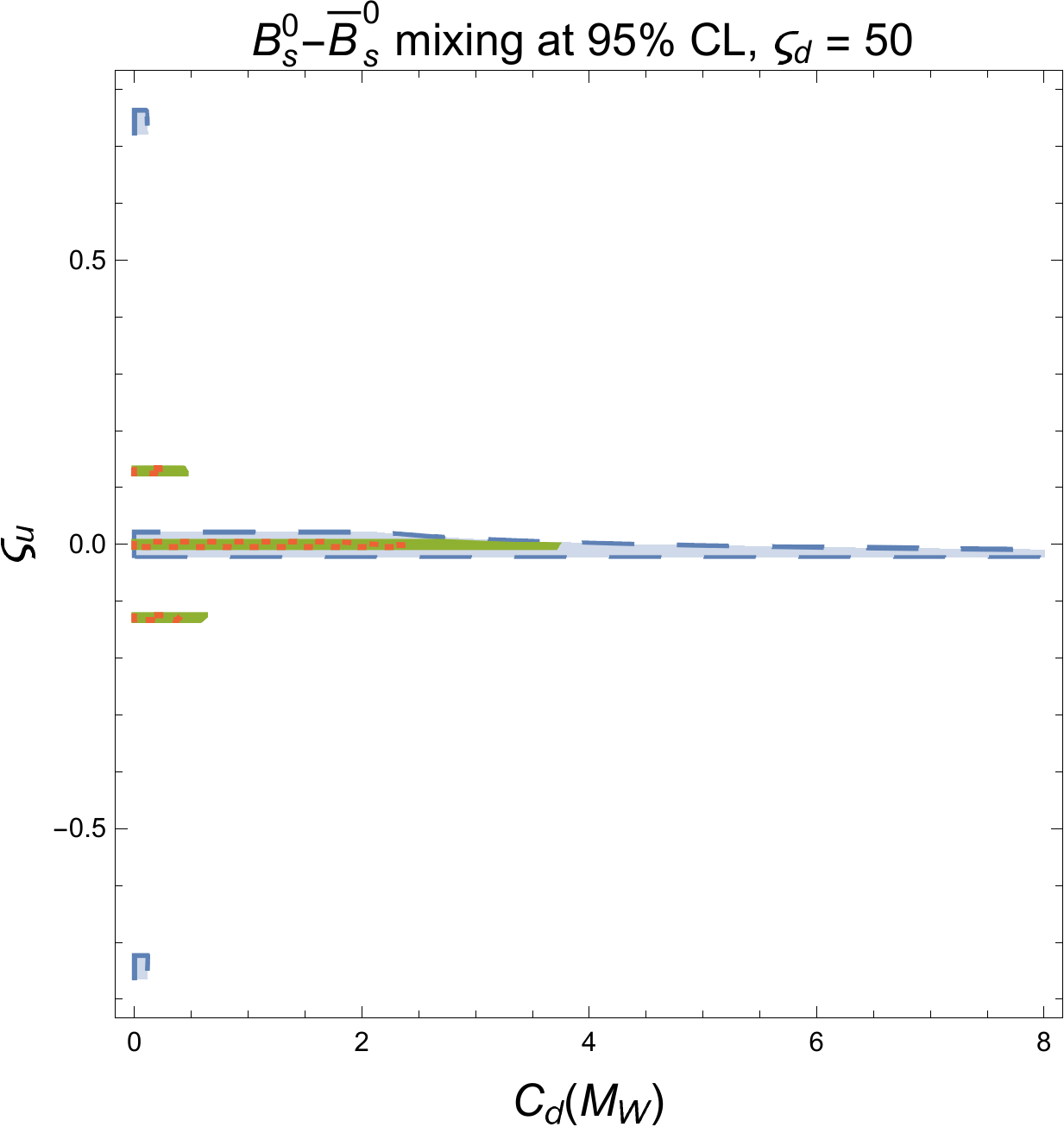}
}
\hskip 1.75cm
\subfloat{
\includegraphics[scale=0.5]{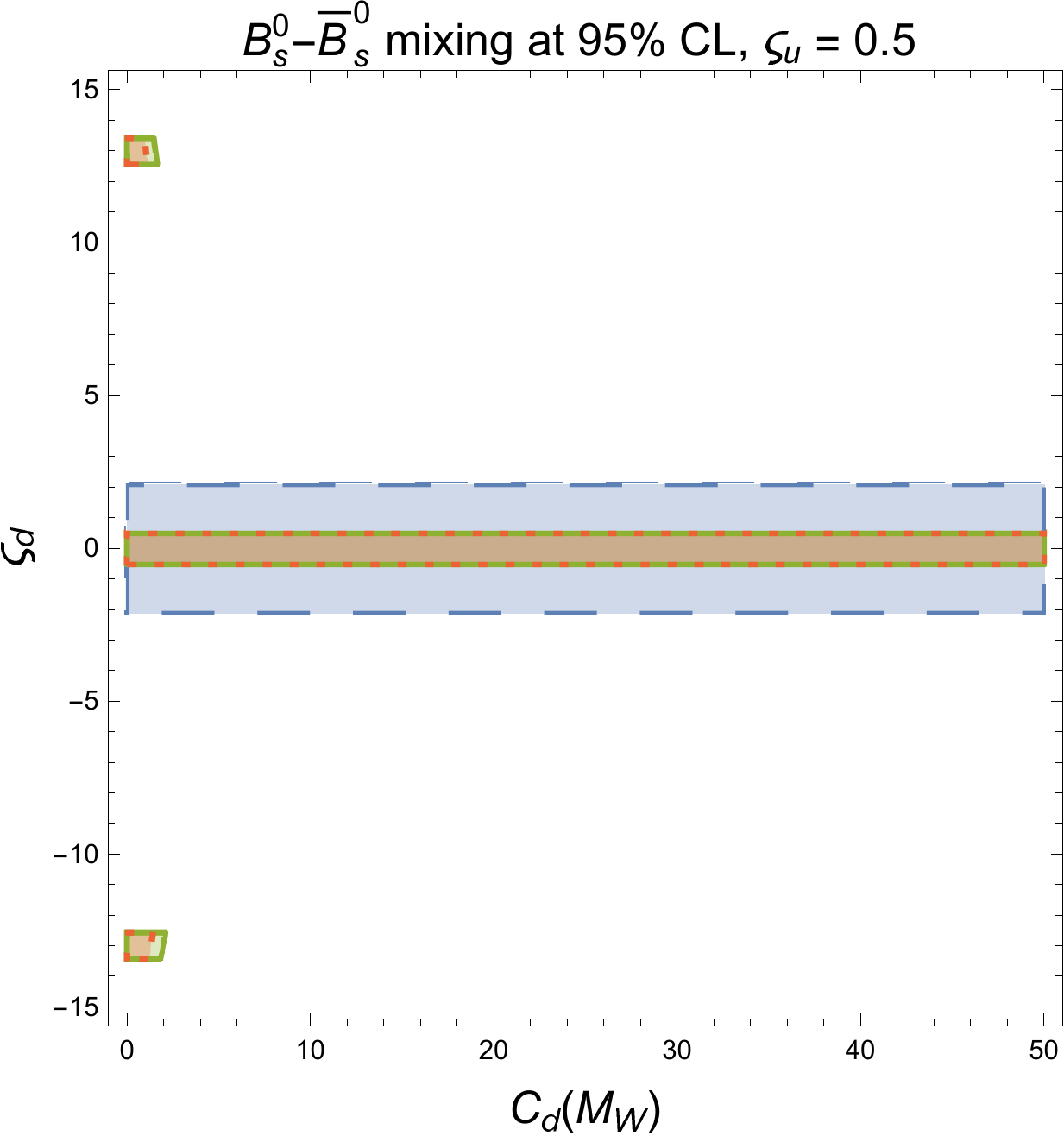}
}
\caption{The left (right) panel shows the $B_s^0$--$\bar B_s^0$ mixing constraints on $C_d(M_W)$ and $\varsigma_{u}$ ($\varsigma_d$) for a fixed value of $\varsigma_d = 50$ ($\varsigma_u = 0.5$), in the CP-conserving  limit and with different mass configurations: A (red, dotted), B (green, solid line) and C/D (blue, dashed).}
\label{fig:ConstraintsMeson}
\end{figure}

The $\Delta B =2$ amplitudes are independent of the leptonic alignment parameter $\varsigma_\ell$. Therefore, the constraints extracted from the $B_s^0$--$\bar B_s^0$ mixing may become relevant at small values of $\varsigma_\ell$ where the $B_s^0\to\ell^+\ell^-$ limits are somewhat weaker. In Fig.~\ref{fig:ConstraintsMeson2}, we display the $B_s^0$--$\bar B_s^0$ mixing constraints obtained for $\varsigma_u=0$ (left) and $\varsigma_d=0$ (right), to be compared with Figs.~\ref{fig:Bs1} and \ref{fig:Bs2}, respectively. The left panel shows indeed that at $\varsigma_u=\varsigma_\ell=0$ 
(the one-loop charged contributions to the mixing are proportional to $\varsigma_u$ and are thus zero) the mixing constraints on $C_d(M_W)$ are stronger than the limits from $B_s^0\to\ell^+\ell^-$. This may be related to the much better experimental precision on $\Delta m_{B_s^0}$ (0.1\%), compared with the present 22\% relative error of the measured $B_s^0 \rightarrow \mu^+ \mu^-$ branching fraction. At $\varsigma_d=0$, however, the previous constraints on Fig.~\ref{fig:Bs2} are stronger. The dominant
one-loop contribution to $B_s^0 \rightarrow \mu^+ \mu^-$ originates then in 
$\Delta C_{10}^{\mathrm{A2HDM}}\propto |\varsigma_u|^2 $ that puts a quite stringent limit on $|\varsigma_u|$. With $\varsigma_d=0$ and $\varsigma_u$ small, the $B_s^0$--$\bar B_s^0$ mixing amplitude becomes insensitive to $C_d(M_W)$, while $B_s^0 \rightarrow \mu^+ \mu^-$ can still constrain this parameter at large values of 
$\varsigma_\ell$.

\begin{figure}
\centering
\subfloat{
\includegraphics[scale=0.5]{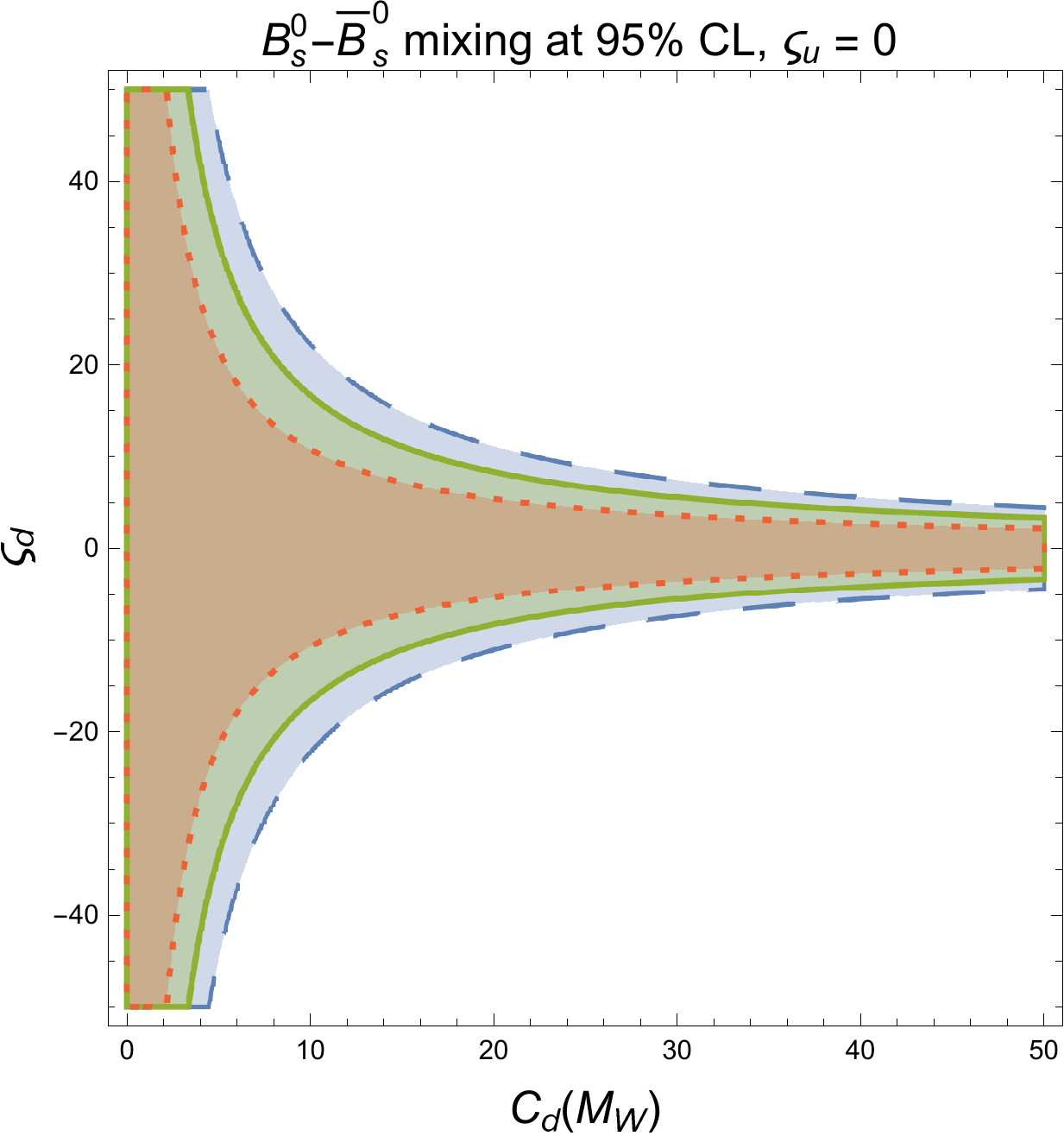}
}
\hskip 1.75cm
\subfloat{
\includegraphics[scale=0.5]{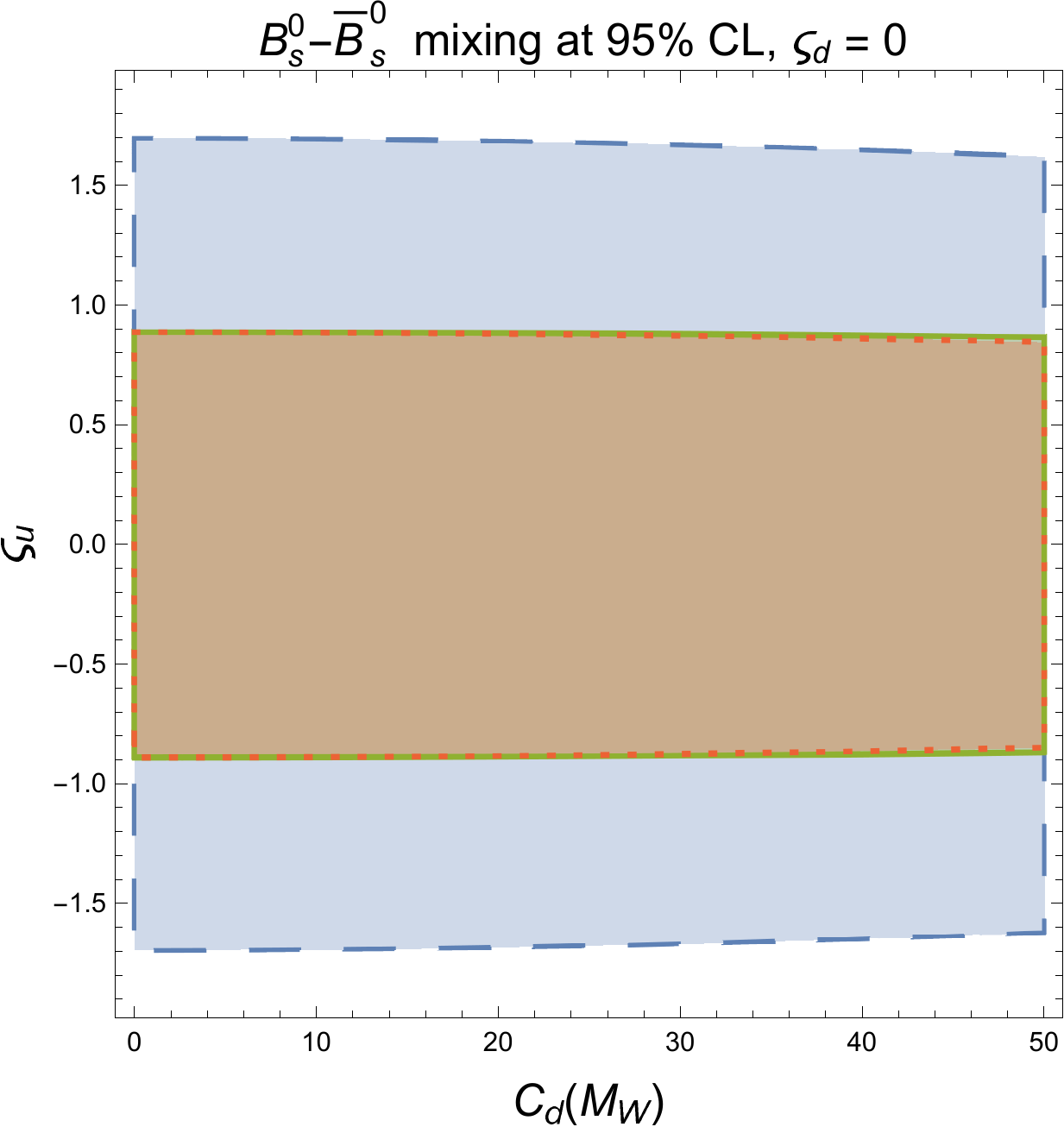}
}
\caption{The left (right) panel shows the $B_s^0$--$\bar B_s^0$ mixing constraints on $C_d(M_W)$ and $\varsigma_{d}$ ($\varsigma_u$) for a fixed value of $\varsigma_u = 0$ ($\varsigma_d = 0$), in the CP-conserving  limit and with different mass configurations: A (red, dotted), B (green, solid line) and C/D (blue, dashed).} 
\label{fig:ConstraintsMeson2}
\end{figure}

\section{Summary}
\label{sec:summary}

The simplicity and versatility of multi-Higgs-doublet models make them favourable candidates for building alternative scenarios of EWSB with extended scalar sectors. The physical spectrum of these models contains a rich variety of bosonic states, with $N-1$ charged and $2N-1$ neutral scalars. The neutral scalar fields can, in general, couple to fermions through non-diagonal flavour interactions, generating unwanted FCNC transitions at tree level that need to be strongly suppressed in order to satisfy the stringent experimental constraints.

One could force these FCNC effects to be unobservable through very small Yukawa couplings or very large scalar masses, making these models irrelevant for present experiments. A more interesting possibility, allowing for new scalar particles not too far from the electroweak scale, is a highly non-generic set of Yukawa couplings. The huge $SU(3)^5$ flavour symmetry of the electroweak Lagrangian is only broken by the Yukawa interactions, but the data clearly indicate that this symmetry breaking only occurs along very specific directions in the flavour space \cite{Chivukula:1987py,DAmbrosio:2002vsn}.

The simplest way to avoid tree-level FCNCs is minimizing drastically the number of flavour couplings, imposing most of them to be zero. Usually, only one scalar doublet is allowed to have Yukawa interactions with a given type of right-handed fermion, fixing in this way a unique flavour-breaking structure associated with each $f'_R$ field. Since this requirement can be always imposed through discrete $\mathcal{Z}_2^{d} \otimes \mathcal{Z}_2^{u} \otimes  \mathcal{Z}_2^{\ell}$ symmetries, the resulting flavour configuration is stable under quantum corrections, leading to the so-called models with natural flavour conservation \cite{Glashow:1976nt,Paschos:1976ay}. With $N > 3$ Higgs doublets, this type of models necessarily involves a minimum of $N-3$ scalar doublets that are decoupled from the fermion sector.

The more general assumption of flavour alignment \cite{Pich:2009sp,Pich:2010ic} is based on the simultaneous diagonalization of all the Yukawa matrices in the fermion-mass eigenstate basis. This implies the appearance of $3(N-1)$ alignment factors, which in the most general case are $3 \times 3$ complex diagonal matrices. In the absence of a specific symmetry protection, the resulting flavour structure is unstable under quantum corrections, which misalign the different Yukawa matrices. Nevertheless, the induced misalignment is a quite small effect, thanks to the residual flavour symmetries of the aligned multi-Higgs Lagrangian, which tightly constrain the type of FCNC operators that can be generated at higher orders.

In this paper, we have studied the misalignment local structure $\cL_{\mathrm{FCNC}}$ induced at one loop, for the most generic aligned multi-Higgs Lagrangian, using the known RGEs of these models. We have particularized the result to different scenarios of phenomenological relevance and have discussed in detail the role of the underlying flavour-dependent phase symmetries. While the misalignment is a very small effect, being suppressed by at least two insertions of the CKM matrix, three Yukawa couplings and the one-loop $1/(4\pi)^2$ factor, it could still lead to interesting phenomenological effects through $V_{tb}^{\phantom{*}} V_{ts}^*m_t^2 m_b$ contributions to effective $\varphi_k^0 \bar s_L b_R$ vertices.

We have investigated the current constraints on the misalignment parameter $C_d(M_W)$, emerging from the measured $B_s^0\to\ell^+\ell^-$ branching fraction and $B_s^0$--$\bar B_s^0$ mixing, taking into account the strong correlated limits on $\varsigma_u$, $\varsigma_d$ and $M_{H^\pm}$ from $\bar B\to X_s\gamma$. These FCNC transitions receive non-local one-loop contributions with internal top and $H^\pm$ propagators \cite{Jung:2010ik,Li:2014fea} that dominate in large regions of the parameter space and were neglected in previous phenomenological studies of the flavour misalignment \cite{Braeuninger:2010td,Bijnens:2011gd,Gori:2017qwg}. The local misalignment Lagrangian $\cL_{\mathrm{FCNC}}$ contributes to these processes through tree-level neutral scalar exchange. For $B_s^0\to\ell^+\ell^-$, where only one insertion of $\cL_{\mathrm{FCNC}}$ is needed, this contribution is actually needed to renormalize the effective $\varphi_k^0 \bar s_L b_R$ vertex and, therefore, appears at the one-loop level. The contribution to $B_s^0$--$\bar B_s^0$ mixing involves, however, two insertions of $\cL_{\mathrm{FCNC}}$; it is a two-loop effect that should be considered together with two-loop diagrams involving two one-loop effective $\varphi_k^0 \bar s_L b_R$ vertices.
We have nevertheless analysed whether the neutral-scalar-exchange amplitude could lead to relevant phenomenological signals through very large values of $C_d(M_W)$.

The present phenomenological constraints on $C_d(M_W)$ are shown in Figs.~\ref{fig:Bs1} to \ref{fig:ConstraintsMeson2}, with different choices of $\varsigma_{u,d,\ell}$ and several benchmark configurations for the scalar mass spectrum. To simplify the analysis we have assumed the absence of any CP-violation effects beyond the usual CKM phase. While stringent bounds emerge on the alignment parameters $\varsigma_{u,d,\ell}$, the sensitivity to $C_d(M_W)$ is very small, as expected, exhibiting the strong phenomenological suppression of the misalignment. 
The local $\cL_{\mathrm{FCNC}}$ contribution is proportional to the product $(\varsigma_u-\varsigma_d) (1+\varsigma_u\varsigma_d) C_d(M_W)$, which explains the pattern displayed by the obtained constraints. Only at large values of $\varsigma_d$ and/or $\varsigma_\ell$ ($|\varsigma_u|$ is bounded to be small) one obtains a somewhat enhanced misalignment contribution that can result in useful limits on $C_d(M_W)$.

The hypothesis of flavour alignment at a very high scale $\mu=\Lambda_A$, {\it i.e.}, $C_{d,u}(\Lambda_A)=0$, survives the phenomenological limits in all cases. With $\Lambda_A \le M_{\mathrm{Planck}} \sim 10^{19}\:\mathrm{GeV}$, it implies $\mathcal{C}_{d,u}(M_W) = \log \frac{\Lambda_A}{M_W} \le 40$, which can easily satisfy all present constraints. This simple relation between $\mathcal{C}_{d,u}(M_W)$ and $\Lambda_A$ has been obtained at the lowest perturbative order. For very large values of the Yukawa couplings and $\Lambda_A\gg M_W$, the long running between the scales $\Lambda_A$ and $M_W$ makes necessary to perform a  resummation of large logarithmic corrections, through a numerical solution of the RGEs
\cite{Braeuninger:2010td,Bijnens:2011gd,Gori:2017qwg} that can modify the high-scale relation by a factor of $\cO(1)$. 
While this slightly changes the scale $\Lambda_A$ associated with a given value of $C_{d,u}(M_W)$, it does not modify our conclusion that high-scale alignment is compatible with all known experimental constraints.

Our phenomenological analyses have been restricted to the simplest case of the A2HDM. Since this is the most constrained scenario of multi-Higgs flavour alignment (the one with the smallest number of free parameters), our conclusion is  obviously also valid for more generic situations with $N>2$ Higgs doublets and/or generalized alignment structures.

\section*{Acknowledgements}
This work has been supported in part by the Spanish Government and ERDF funds from
the EU Commission [Grant FPA2014-53631-C2-1-P], and by the Spanish
Centro de Excelencia Severo Ochoa Programme [Grant SEV-2014-0398].
The work of Ana Peñuelas is funded by Ministerio de Educación, Cultura y Deporte, Spain [Grant FPU15/05103].

\addappheadtotoc
\appendix

\section{Hadronic matrix elements for meson mixing}
\label{app:Bpar}

The Wilson coefficients of the effective Hamiltonian~\eqn{eq:Heff} have been evaluated at the electroweak scale, 
$\mu_{tW} \sim \cO(M_W, m_t,M_{H^\pm},M_{\varphi_i^0})$, 
and need to be evolved down to the low-energy scales where the hadronic matrix elements of the corresponding quark operators are determined. In addition to the three scalar operators in Eq.~\eqn{eq:DF2op}, generated through $\varphi^0_k$-exchange between two $\cL_{\mathrm{FCNC}}$ vertices, one must take also into account the leading contributions from 1-loop box diagrams with $W^\pm$ and/or $H^\pm$ propagators. Neglecting the light quark mass ($m_{d,s}$ for $B_{d,s}^0$ or $m_d$ for $K^0$), these charged-current boxes contribute to $C_{1,ij}^{SRR}$ and to the SM operator \cite{Jung:2010ik}
\beq
\cO^{VLL}_{ij}\, =\, (\bar{d}_{i L} \gamma_\mu d_{j L}) (\bar{d}_{i L} \gamma^\mu d_{j L})\, .
\eeq
Gluonic corrections give rise to the appearance of additional operators which mix under renormalization with the previous ones. In general, one must consider a basis of eight operators including the additional structures~\cite{Buras:2000if}:
\beqn\label{eq:DF2opadd}
\cO^{VRR}_{ij}\, =\, (\bar{d}_{i R} \gamma_\mu d_{j R}) (\bar{d}_{i R} \gamma^\mu d_{j R})\, ,
\qquad\qquad
\mathcal{O}_{1,ij}^{LR} \, =\, (\bar{d}_{i L} \gamma_\mu d_{j L}) (\bar{d}_{i R} \gamma^\mu d_{j R}) \, , 
\no\\[5pt]
\mathcal{O}_{2,ij}^{SLL} \, =\, (\bar{d}_{i R} \sigma_{\mu\nu} d_{j L}) (\bar{d}_{i R}  \sigma^{\mu\nu} d_{j L})\, ,  
\qquad\qquad
\mathcal{O}_{2,ij}^{SRR} \, =\, (\bar{d}_{i L} \sigma_{\mu\nu} d_{j R}) (\bar{d}_{i L}  \sigma^{\mu\nu} d_{j R})\, ,  
\eeqn
with\footnote{Notice that Refs.~\cite{Buras:2000if,Buras:2001ra} adopt a non-conventional definition of $\sigma^{\mu\nu}$, without the factor `$i$', and have then the opposite sign for the operators $\mathcal{O}_{2,ij}^{SRR}$.} $\sigma^{\mu\nu} \equiv \frac{i}{2}\left[\gamma^\mu,\gamma^\nu\right]$.
The renormalization group evolution of this operator basis factorizes in five different sectors~\cite{Buras:2000if,Buras:2001ra}:
\begin{equation}
\begin{bmatrix}
C_{1, ij}^{X} (\mu) \\[8pt]
C_{2, ij}^{X} (\mu) 
\end{bmatrix}\; =\; \begin{bmatrix}
\left[ \eta_{11}(\mu) \right]_{X}  & \left[ \eta_{12}(\mu) \right]_{X}  \\[8pt]
\left[ \eta_{21}(\mu) \right]_{X}  & \left[ \eta_{22}(\mu) \right]_{X} 
\end{bmatrix} \; \begin{bmatrix}
C_{1, ij}^{X} (\mu_{tW}) \\[8pt]
C_{2, ij}^{X} (\mu_{tW}) 
\end{bmatrix}  \, , 
\eeq
\beq
C_{1, ij}^{Y} (\mu)\; =\; \left[ \eta(\mu) \right]_{Y}\;  C_{1, ij}^{Y} (\mu_{tW})  \, ,
\end{equation}
where $X = SRR, SLL, LR$ and $Y= VLL, VRR$.
Next-to-leading-order expressions for the coefficients $\left[ \eta_{kl}(\mu) \right]_{X}$ ($k,l = 1,2$)
and $\left[ \eta(\mu) \right]_{Y}$ can be found in Refs.~\cite{Buras:2000if,Buras:2001ra} for the $B_q^0$ and $K^0$ systems.
Since in our case the initial conditions are only known at the lowest order, we have calculated the evolution with leading-order anomalous dimensions and two-loop running for the strong coupling $\alpha_s$.

The hadronic matrix elements of the $\Delta F=2$ four-quark operators  can be 
expressed as:
\beqn
\bra{\bar{M}^0} \mathcal{O}^{VZZ}_{1, ij} \ket{M^0}  &=& \frac{2}{3}\, f_{M}^2\, m_{M^0}^2\; B_1^{VZZ}(\mu)\, , 
\label{eq:BVLL} \\
\bra{\bar{M}^0} \mathcal{O}^{LR}_{1, ij} \ket{M^0}  &=& - \frac{1}{3}\, \left(\frac{f_{M}\, m_{M^0}^2}{m_i(\mu) + m_j(\mu)}\right)^2\; B_1^{LR}(\mu) \, ,
\label{eq:LR1} \\
\bra{\bar{M}^0} \mathcal{O}^{LR}_{2, ij} \ket{M^0}  &=& \frac{1}{2}\, \left(\frac{f_{M}\, m_{M^0}^2}{m_i(\mu) + m_j(\mu)}\right)^2\; B_2^{LR}(\mu) \, ,
\label{eq:LR2} \\
\bra{\bar{M}^0} \mathcal{O}^{SZZ}_{1, ij} \ket{M^0}  &=&  -\frac{5}{12} \left(\frac{f_{M}\, m_{M^0}^2}{m_i(\mu) + m_j(\mu)}\right)^2\; B_1^{SZZ} (\mu) \, ,
\label{eq:BSLL1} \\
\bra{\bar{M}^0} \mathcal{O}^{SZZ}_{2, ij} \ket{M^0}  &=&   \left(\frac{f_{M}\, m_{M^0}^2}{m_i(\mu) + m_j(\mu)}\right)^2\; B_2^{SZZ} (\mu) \, , 
\label{eq:BSLL2}
\eeqn
where $Z=L,R$ denotes the two different operator chiralities, $m_{i,j}(\mu)$ are the relevant running quark masses and the $B_i(\mu)$ factors parametrize the deviations from the naive vacuum-insertion approximation.
%
These parameters have been calculated by the ETM lattice collaboration, employing the ratio method approach on $N_f=2$ ensembles for $B_{d}^0$ and $B_{s}^0$ \cite{Carrasco:2013zta}, and simulations with $N_f = 2 + 1 + 1$ dynamical sea quarks for $K^0$ \cite{Carrasco:2015pra}.  The ETM results are given in a different operator basis; the connection reads:
\beq
B_1^{VZZ}(\mu)\, =\,  B_1(\mu)\, ,  
\qquad\qquad
B_1^{LR}(\mu)\, =\, B_5(\mu) \, ,   
\qquad\qquad
B_2^{LR}(\mu)\, =\, B_4(\mu)  \, , 
\no\eeq\\[-30pt]
\beq   
B_1^{SZZ}(\mu)\, =\, B_2(\mu) \, ,    
\qquad\qquad
B_2^{SZZ}(\mu)\, =\,  \frac{5}{3}\, B_2(\mu)- \frac{2}{3}\,  B_3(\mu)   \, .
\eeq
The numerical values of the $B_i$ parameters are compiled in Table~\ref{table:Bparameters}.

\begin{table}[t]
\centering
{\renewcommand{\arraystretch}{1.3}
\resizebox{\textwidth}{!}{%
\begin{tabular}{|c |c  c c c c |} 
 \hline
 $i =$ & 1 & 2 & 3 & 4 & 5  \\ 
 \hline\hline
$ f_{B_d}\sqrt{B_{i}^{B_d}}$ & $ 174 \pm 8 $ MeV & $160 \pm 8$ MeV & $177 \pm 17$ MeV & $185 \pm 9$ MeV  & $229 \pm 14$ MeV \\
$ f_{B_s}\sqrt{B_{i}^{B_s}}$ & $211 \pm 8$ MeV &  $195 \pm 7$ MeV & $215 \pm 17$MeV & $220 \pm 9$ MeV & $285 \pm 14 $ MeV \\
$ B_{i}^{K}$ & $0.506 \pm 0.017 \pm 0.003$ & $ 0.46 \pm 0.01 \pm 0.03 $ & $ 0.79 \pm 0.02 \pm 0.05 $ & $ 0.78 \pm 0.02 \pm 0.04$ & $0.49 \pm 0.03 \pm 0.03  $ \\
 \hline
  \end{tabular}
  }
  }
\caption{Lattice determinations of $f_{M} \sqrt{B_i^M}$  ($ M=B_d^0, B_s^0$)  \cite{Carrasco:2013zta}  and $B_i^K$ ($M=K^0$)   \cite{Carrasco:2015pra}, in the $\overline{\mathrm{MS}}$ scheme. The $B_q^0$ parameters are given at $\mu= m_b$, while the $K^0$ values refer to $\mu = 3~\mathrm{GeV}$.}
\label{table:Bparameters}
\end{table}

The observables relevant for our phenomenological analyses are
\begin{equation}
\label{eq:observables}
\Delta m_{B_q^0}\, =\, \frac{1}{m_{B_q^0}}\; \abs{ \bra{B_q^0} \mathcal{H}_{\mathrm{eff}} \ket{\bar{B}_q^0}} \, , 
\qquad\qquad 
\varepsilon_K\, =\, k_{\epsilon} \;\frac{\e^{i \phi_{\epsilon}}}{\sqrt{2}}\; 
\frac{\Im \left( \bra{K^0} \mathcal{H}_{\mathrm{eff}} \ket{\bar{K}^0}\right)}{2 m_K\,\Delta m_K} \, ,
\end{equation}
where $\phi_{\epsilon} \approx \tan^{-1}{[2 (m_{K_L} -m_{K_S})/(\Gamma_{K_S} -\Gamma_{K_L})]} = (43.52\pm 0.05)^\circ$ is the so-called superweak phase \cite{Patrignani:2016xqp} and $k_{\epsilon}\approx 0.94\pm 0.02 $ accounts for small long-distance corrections \cite{Buras:2010pza}. We do not extract new-physics constraints from $\Delta m_K$ because the kaon mass difference receives large long-distance contributions that introduce sizeable theoretical uncertainties.


\end{document}